\documentclass[preprint2]{proto}
\usepackage{times}
\usepackage{graphicx}
\usepackage{natbib}

\begin{document}

\title{\textbf{\LARGE PHYSICAL EFFECTS OF COLLISIONS IN THE KUIPER BELT} }

\author{\textbf{\large ZO\"E M. LEINHARDT}}
\affil{\small \em Harvard University}
\author{\textbf{\large SARAH T. STEWART}}
\affil{\small \em Harvard University}
\author{\textbf{\large PETER H. SCHULTZ}}
\affil{\small \em Brown University}

\begin{abstract}
\noindent Collisions are a major modification process over the history of the
Kuiper Belt. Recent work illuminates the complex array of possible
outcomes of individual collisions onto porous, volatile bodies. The
cumulative effects of such collisions on the surface features, 
composition, and internal structure of Kuiper Belt Objects are not yet
known. In this chapter, we present the current state of knowledge of
the physics of cratering and disruptive collisions in KBO analog
materials. We summarize the evidence for a rich collisional history in 
the Kuiper Belt and present the range possible physical modifications
on individual objects. The question of how well present day bodies
represent primordial planetesimals can be addressed through future
studies of the coupled physical and collisional evolution of Kuiper 
Belt Objects. \vspace{0.5cm}
\end{abstract}

\section{INTRODUCTION}

The Kuiper Belt contains some of the least modified material in the
solar system. Some Kuiper Belt objects (KBOs) may be similar to the
planetesimals that accreted into the larger bodies in the outer solar
system. However, KBOs have suffered various modification processes
over the lifetime of the solar system, including damage from cosmic
rays and ultraviolet radiation, sputtering and erosion by gas and dust
in the interstellar medium, and mutual collisions
\citep[e.g.][]{Stern03}.  Robust interpretations of the surfaces and
internal structures of KBOs require improved insight into the relative
weight of each of these processes.

The present understanding of the importance of collisions on the
physical evolution of KBOs is limited by the state of knowledge in
three fundamental areas: (1) the dynamical history of the different
populations within the trans-Neptunian region ({\em Morbidelli et al.}
this volume); (2) the physical properties of KBOs ({\em Brown}, {\em
  Stansberry et al.} this volume); and (3) how the physical properties
of KBOs (expected to be icy and porous) affect the outcome of
collisions ({\em this chapter}). The dynamical history of a population
defines the evolution of mean impact parameters (velocity, angle, mass
ratio of the projectile and the target) within and between KBO
populations. The impact parameters and the material properties of the
colliding bodies determine the outcome of an individual impact event.
Finally, the cumulative effects of collisions are determined by the
coupled physical and dynamical evolution of KBOs.

Variable progress has been made in these three areas. Over the past
decade, great improvements in observations and models have illuminated
the rich dynamical history of the Kuiper Belt. At present, there is a
sparse, but growing, body of data on the physical properties of KBOs
(e.g., size, density, composition, and internal structure).  Although
a significant body of work has been devoted to collisions between icy,
porous bodies, our understanding of the governing physics is still
incomplete.
%A detailed model of the collision itself is needed to
%understand the coupled physical and dynamical history of KBOs.
The collisional evolution of KBOs is a particularly interesting and
challenging problem because of the range of possible outcomes that
depend on the changing dynamical structure of the Kuiper Belt.

In this chapter, we present a summary of the work to date that can be
applied to the physical effects of collisions in the Kuiper Belt.  We
begin with observational evidence for significant past and present-day
collisions in the Kuiper Belt (\S \ref{sec:obs}).  We then present a
range of possible outcomes from collisions between KBOs (\S
\ref{sec:poss}) and discuss the principal discriminating factors (\S
\ref{sec:fact}).  Based on the expected physical properties of KBOs,
we summarize the results of laboratory and numerical experiments that
have been conducted to determine how material properties, such as
composition, porosity and impact conditions, including velocity and mass ratio, affect collision outcomes (\S
\ref{sec:studies}). Finally, we discuss several open questions and
future research directions for studying collisions in the Kuiper Belt. (\S
\ref{sec:questions}).

\section{EVIDENCE FOR A RICH COLLISIONAL HISTORY IN THE KUIPER BELT}\label{sec:obs}

In this section, we summarize four observations that support a
significant collisional history within the Kuiper Belt.  First, we
discuss observations of interplanetary dust particles (IDPs) by the
Pioneer and Voyager spacecraft (\S \ref{sec:dust}).  Analyses of the
orbits of IDPs conclude that the Kuiper Belt must be one of the dust
source regions. Second, the size distribution of Kuiper Belt Objects
has at least one break from a simple power law around diameters of
10's km, which is consistent with models of collisional equilibrium
among the smaller bodies (\S \ref{sec:size}). Third, the discovery of
a possible dynamical family of objects in the Kuiper Belt implies
conditions that produced at least one near-catastrophic collision of
one of the largest KBOs (\S \ref{sec:fam}). Finally, models of the
accretion of the largest KBOs demonstrate that the mass in the ancient
Kuiper Belt must have been much larger than observed today. The total
mass loss of $>$90\%, and perhaps as much as 99.9\%, was driven by a
combination of dynamical perturbations and collisional grinding (\S
\ref{sec:mass}).

In addition to the observable features discussed below, collisions
within a small body population will also affect rotation rates,
surface colors, and the formation of binaries. The rotation rates of
bodies in collisional equilibrium will reflect the angular momentum
transfer from typical impact conditions \citep[see e.g.,][for asteroid
rotations]{Love97, Paolicchi02}.  The formation of binary KBOs
is still a matter of debate. Some binaries seem to have formed via
collisions, while others have too much angular momentum for a
collision origin \citep{Margot02}.  The observed color diversity in
the Kuiper Belt is also controversial and not correlated directly with collision energy (see {\em Doressoundiram et al.} this volume).
However, the range of outcomes from collisions depend on material
properties as well as the impact parameters. The growing data on
rotation rates, colors, and binaries will provide in the future additional
constraints on the collisional evolution in the Kuiper Belt.

\subsection{Interplanetary Dust Particles}\label{sec:dust} 

Dust and small meteoroids were detected in the outer solar system by
the Pioneer 10 \& 11 and Voyager 1 \& 2 missions
\citep{Humes80,Gurnett97}.  Pioneer 10 and 11 measured the
concentration and orbital properties of dust from 1 to 18 AU. The dust
impacts detected by Pioneer 11 between 4 and 5 AU were determined to
have either high inclination or eccentricity or both. In other words,
the IDPs were either not on circular orbits and/or not on near planar
orbits. Hence, the observed increase in particle flux at Jupiter could
not be explained by gravitational focusing, which is inefficient for
highly inclined and eccentric orbits, and \citet{Humes80} suggested
that the dust had a cometary origin.

In a reanalysis of the Pioneer data, \citet{Landgraf02} found that the
dust flux was relatively constant at distances exterior to Jupiter's
orbit. To produce a constant dust flux from drag forces, the dust must
originate from a source beyond the detection locations by the
spacecraft.  \citet{Landgraf02} modeled the radial dust contribution
using three source reservoirs, dust from evaporating Oort Cloud and
Jupiter family comets and dust from collisions between KBOs, and argue
that the amount of dust observed by Pioneer 10 and 11 can only be
explained by a combination of all three reservoirs.  They find that
comets can account for the material detected inside Saturn's orbit but
an additional reservoir is necessary for the dust observed
outside Saturn's orbit.

Although Voyager 1 and 2 did not carry specialized detectors for dust,
\citet{Gurnett97} found that the plasma wave instruments could detect
impacts from small particles with masses $\ge 10^{-11}$~g (two to
three orders of magnitude below the mass detection limit by Pioneer 10
and 11). From data collected between 6 and 60 AU, \citet{Gurnett97}
found a severe drop off in dust detection events after 51 AU and 33 AU
for Voyager 1 and 2, respectively. As a result, the authors conclude
that the source of the dust cannot be interstellar. Furthermore, the
small latitudinal gradient decreases the likelihood that the source
objects are planets, moons, or asteroids (if the dust did originate
from such objects, one would expect a strong latitudinal gradient
since the planets, moons, and asteroids are effectively all in the
same plane). The Voyager IDP observations are consistent with a dust
source from the Kuiper Belt \citep{Gurnett97,Jewitt00}.

In summary, the radial distribution and orbital properties of outer
solar system IDPs cannot be explained by source material solely from
Jupiter family comets and Oort cloud comets and indicate the need for
an additional active source of dust in the outer solar system.  The
IDP observations are well matched by models of dust produced during
the collisional evolution of the Kuiper Belt \citep[e.g.,][and \S
\ref{sec:mass}]{Jewitt00}. Dust derived from mutual collisions in the present
day Kuiper Belt is analogous to the zodiacal dust from the asteroid
belt \citep{Muller05} and observations of rings of dust around other main
sequence stars ({\it Moro-Martin et al.} and {\it Liou et al.} this volume).  Because the
removal time of dust is much shorter than the age of the solar system
({\em Kenyon et al.} this volume), the dust must be replenished by collisions occurring
throughout the history of the solar system.

\subsection{Size Distribution of KBOs}\label{sec:size}

%{\bf have either of us read Kenyon's chapter yet?}

Formation models indicate that KBOs accreted within a thin disk with
low relative velocities and inclinations ({\em Morbidelli et al.}
this volume).  However, the present velocity dispersion ($\sim 1$ km
s$^{-1}$) and the inclination distribution (about 20 degree half
width) of KBOs are both much higher than
expected during the coagulation stage \citep{Trujillo01}.  The large
relative velocities and the large inclination distribution of the KBOs
point to significant dynamical interactions with Neptune, which
resulted in a rich collisional history \citep{Davis97}. 
%{\bf I feel
 % like this is a weak opening because there are lots of observations
  %(e.g., resonances) that point to interactions with Neptune and
  %collisional history.}

%The dynamical history of the Kuiper Belt has led to the development of
%four distinct populations ({\em Gladman et al.} this volume): 1) {\em
%  Classical Kuiper Belt Objects (CKBOs)} have relatively low
%eccentricity $< 0.24$ and $41 < a < 2000$ AU protecting them from
%current interactions with Neptune; 2) {\em Resonant Kuiper Belt
 % Objects} are located in mean motion resonances with Neptune which
%means that they are also protected from direct interaction with
%Neptune; 3) {\em Scattering Kuiper Belt Objects} are currently
%scattering off of Neptune with semi-major axes $> 30$ AU; and 4) {\em
%  Detached Kuiper Belt Objects} are not currently scattering off of
%Neptune and have high eccentricities $e > 0.24$. The size
%distribution, surface features and colors, composition, and internal
%structures of bodies within each population may reflect differences in
%solar distance during accretion and dynamical histories.

If the bodies in the Kuiper Belt were fully collisionally evolved and
collision outcomes were independent of size, the differential size
distribution ($dN \sim r^{-q}dr$, where $N$ is number of bodies in the
size bin of radius $r$) would be described by a self-similar
collisional cascade and fit by a single power law index of $q=3.5$
\citep{Dohnanyi69,williams94}.  If the population is only partially
collisionally evolved and/or the disruption criteria is dependent on
scale, the size distribution will deviate from a single power law.
For example, the size distribution in the asteroid belt deviates from
a simple power law in part because of strength effects
\citep{Obrien05} and recent collisions, such as dynamical family
forming events \citep{Delloro01b}.

Recent observations indicate that the size distribution of KBOs has a
break at diameters of 10's km, with fewer smaller bodies than expected
from extrapolation from bodies of 100's km diameter
\citep{Bernstein04,Chen06,Roques06}. The slope of the differential
size distribution of large KBOs ($<$ 25 magnitude, $>$100 km diameter)
is well established, with a slope in the range of 4 to 4.8
\citep[][{\em Petit et al.} this volume]{Trujillo01,petit06}. \citet{Bernstein04} also suggested that
the classical KBOs have a different size distribution from the other
dynamical populations (for KBO population classifications, see {\em
  Gladman et al.} this volume).

%\citet{Bernstein04} used the size
%distribution from \citet{Trujillo01} to predict that they should
%detect 84 new Kuiper Belt Objects with a deeper survey using the
%Advanced Camera for Surveys on the Hubble Space Telescope. However,
%\citet{Bernstein04} found only three new KBOs. These surprising
%results meant a significant deviation from the single power-law fit to
%surface density and size distribution.  Combining the results of the
%ACS survey with results from past surveys, \citet{Bernstein04} find
%significant departures from the single power-law fit to surface
%density for both large and small objects. 
%
%In addition, fits to
%different dynamical classes within the Kuiper Belt suggest with high
%statistical significance that classical and excited populations have
%different size distributions. The excited population seems to have
%more mass in larger objects. This suggests that the excited objects
%seem to have grown to larger sizes. One explanation is that the
%excited objects were originally interior to the classical objects
%where both the collision rates and the surface density was larger than
%the outer region accelerating the accretion timescale.
% MAYBE ADD THIS BACK IN SOMEWHERE ELSE?

Over the past decade, several groups have made significant progress in
modeling the collisional evolution of the Kuiper Belt \citep[][{\em
  Kenyon et al.} this
volume]{Davis97,Stern97,Kenyon04,Pan05,Kenyon99}.  Their work provides
a theoretical basis for a break in the KBO size distribution around
10's km. \citet{Davis97} first demonstrated that few of the largest
bodies in the Kuiper Belt experience catastrophic disruption events in
which 50\% of the mass is permanently removed. In other words, most of
the largest KBOs are primordial; they have persisted since the end of
the coagulation stage, although some may have suffered shattering
collisions.

Collision evolution models indicate that the break in the size
distribution corresponds to the upper size limit of the collisionally
evolved population \citep{Davis97,Kenyon04,Pan05}.  Over time,
collisions preferentially disrupt smaller objects because of their
lower critical disruption energies and their higher number densities
(and hence higher collision probabilities) compared to larger bodies.
When the disruption criteria is size-dependent, the collisionally
evolved size distribution deviates from $q=3.5$
\citep[see][]{obrien03}.  For example, \citet{Pan05} utilize a
disruption criteria proportional to the gravitational binding energy
of the body, and the equilibrium power law has $q=3$.  Numerical
evolution simulations by \citet{Kenyon04} and analytical work by
\citet{Pan05} are in good agreement with the observation of the number
of 10's km size bodies by \citet{Bernstein04}.  Note that the location
of the break in size between the collisionally evolved and primordial
populations increases with time and depends on the dynamical evolution
of the system.

%\citet{Pan05} predict the transition size between collisionally
%evolved and primordial populations to occur at about 40 km radius,
%assuming that the velocity dispersion and mass of the Kuiper Belt have
%not changed significantly over the age of the solar system.  However,
%if the velocity dispersion remained at the low levels expected during
%accretion (e.g., 1 m s$^{-1}$) for an extended period (e.g., 3 Gyr),
%the break in slope of the size distribution should occur at 20 km
%radius, depending on when the velocity dispersion transitioned to the
%present km s$^{-1}$ regime.

%\citet{Chang06} used the Proportional Counter Array on the Rossi X-ray
%Timing Explorer to look for KBOs occulting the bright X-ray source
%Scorpius X-1. The authors found 58 detections that they attribute to
%occultations by small bodies with orbits beyond Neptune. By assuming a
%mean distance for the small bodies of 43 AU, the authors determined
%that size the occulting objects were 50~m on average and up to 100 m
%diameter.  Extrapolation of the single power-law size distribution
%from \citet{Trujillo01} is consistent with the findings of
%\citet{Chang06}.  These observations are not consistent with
%\citet{Bernstein04} or \citet{Pan05}. 

The size distribution of KBOs is likely to have a second break in
slope at significantly smaller sizes. The second break corresponds to
the transition between the collisionally evolved strength-dominated
bodies and collisionally evolved gravity-dominated bodies. As the
strength of KBOs is essentially unknown at present, the location of
the strength to gravity transition based on catastrophic disruption
models, e.g., 10's to 100's~m (\S \ref{sec:cat}), is highly uncertain.

% e.g., $\sim$100~m to 1~km \citep{Benz99,Weidenschilling97}, is
%{\bf any reason to city Benz and Weidenshililng here? what did
 % Weidenshiling do anyway?}

%However, if the break
%occurs at around 100~m and the power-law slope follows the Dohnanyi
%slope of $q=3.5$, the X-ray occulation observations are still
%overabundant \citep{Pan05}.

In summary, the observed size distribution of KBOs departs from a
single power law, which is consistent with the existence of both a
collisionally evolved population and a primordial population
\citep{Davis97,Kenyon04,Pan05,Stern97}. In this scenario, the
primordial populations should have experienced primarily surface
modification processes through impact cratering events
(\S \ref{sec:poss}, Figs.~\ref{fig:color} and \ref{fig:cartoon_evol}ab). The population
of largest bodies probably has a subpopulation with differentiated
internal structures \citep{Merk06} and a subpopulation with rubble
pile structures \citep{Davis97}. Both the collisionally evolved
strength and gravity-dominated populations would have experienced the
full range of collisional outcomes, including catastrophic disruption
and changes in internal structure and composition (\S \ref{sec:poss}, Fig.
\ref{fig:cartoon_evol}c).

\subsection{Kuiper Belt Family (2003 EL61)}\label{sec:fam}

At present, a few tens of dynamical families have been identified in
the asteroid belt \citep{Bendjoya02}. These objects are grouped
together in proper element space and have similar spectral features
where detailed observations are available. The orbits of the family
members can be integrated back in time to a common starting point,
suggesting formation via a catastrophic collision. Although
collisional evolution models of the Kuiper Belt (\S \ref{sec:size})
indicate that very few KBOs larger than about $100$~km have
experienced catastrophic disruption events, \citet{Brown07} have
observed what seems to be a collisional family in the Kuiper Belt (see
also {\em Brown} this volume).

2003 EL61 has two known satellites and five proposed family members.
All of these objects have similar proper elements, colors, and a deep
H$_2$O spectral feature \citep{Brown07}.  Although the detection
of H$_2$O ice on the surfaces of KBOs is not unique to these objects,
the significant depth of the spectral feature is characteristic of the
proposed family, suggesting either more recent or more abundant
exposure of surface ice compared to other KBOs.  2003 EL61 has a
double-peaked rotational light curve with a period of 3.9 hours
\citep{Rabinowitz06}. Assuming that the light curve is solely due to
the equilibrium shape of a rotating, homogeneous, {\it fluid}
ellipsoid, \citet{Rabinowitz06} and \citet{Lacerda06} derive the size
($\sim$1500 km diameter) and density ($\sim 2.6$ g cm$^{-3}$).
However, the derived size and density are highly uncertain as 2003
EL61 is likely to possess nonzero shear strength \citep{Holsapple07b}.

These combined observations suggest that 2003 EL61 suffered a
significant, but sub-catastrophic impact event \citep{Brown07}.  If
the modeled bulk density of 2.6 g cm$^{-3}$ is correct and the
pre-collision density of 2003 EL61 was comparable to other large KBOs
($\sim$2 g cm$^{-3}$, {\em Brown} this volume), then about 20\% of the
original mass was lost.  In this model, the dispersed material was
preferentially H$_2$O ice, presumably derived from an ice-rich mantle,
producing the shared water spectral feature of the proposed family
members.

Further investigation of the proposed 2003 EL61 family and search for
other dynamical families would provide useful constraints on the
collisional history of the Kuiper Belt.

\subsection{Total Mass of the Kuiper Belt}\label{sec:mass}
%{\bf add something about Olber's Paradox? -- see Kenyon}

The total mass in the modern Kuiper Belt is depleted from a smooth
surface density extrapolation from the giant planet region of the
solar system. Based on the observed size distribution of bodies
between 30 to 50 AU, the total mass is only about 0.01 Earth masses
(less than 5 Pluto masses) \citep{Bernstein04}.  However,
\citet{Stern96} and \citet{Stern97} demonstrate that the Kuiper Belt
must have been more massive in the past for the largest KBOs (100 to
1000 km) to accrete via mutual collisions.  At least 90\% of the mass
in the Kuiper Belt was lost through collisions and ejections induced
by the stirring and migrating of Neptune \citep{Stern97,Hahn99}.

\section{POSSIBLE COLLISION OUTCOMES IN THE KUIPER BELT}\label{sec:poss}

\begin{figure}
 \includegraphics[width=80mm]{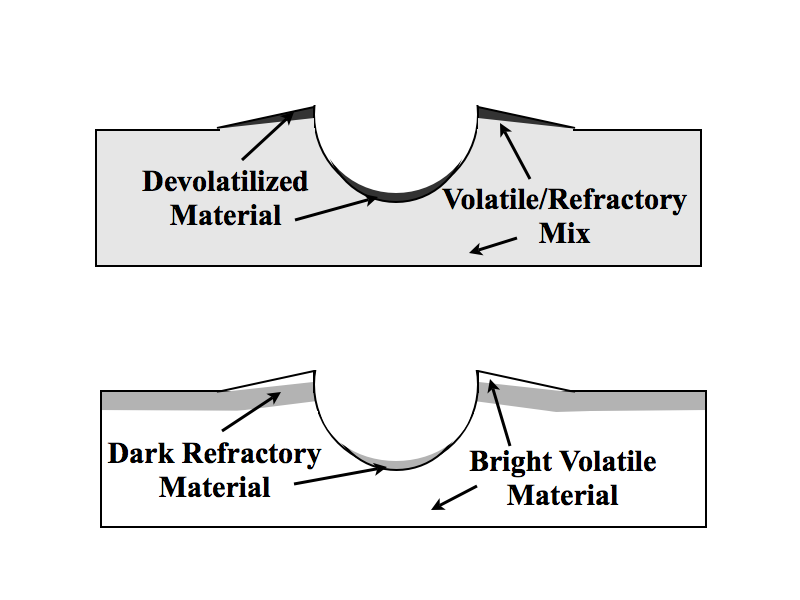} 
 \caption{\small Top: Schematic of an impact crater on a target made of
   a mixture of volatile and refractory material. The energy of the
   impact produces melting and devolatization at the base of the
   crater and in the ejecta. Bottom: Schematic of an impact crater on
   a target made mostly of volatile material. The surface of the
   target is covered with a ``crust'' of darker refractory material.
   The impact is large enough to excavate fresh volatiles from depth
   creating a bright ejecta blanket. Collapse of the crater wall
   creates a darker region at the bottom of the
   crater.\label{fig:color}}
\end{figure}

\begin{figure}
 \includegraphics[width=80mm]{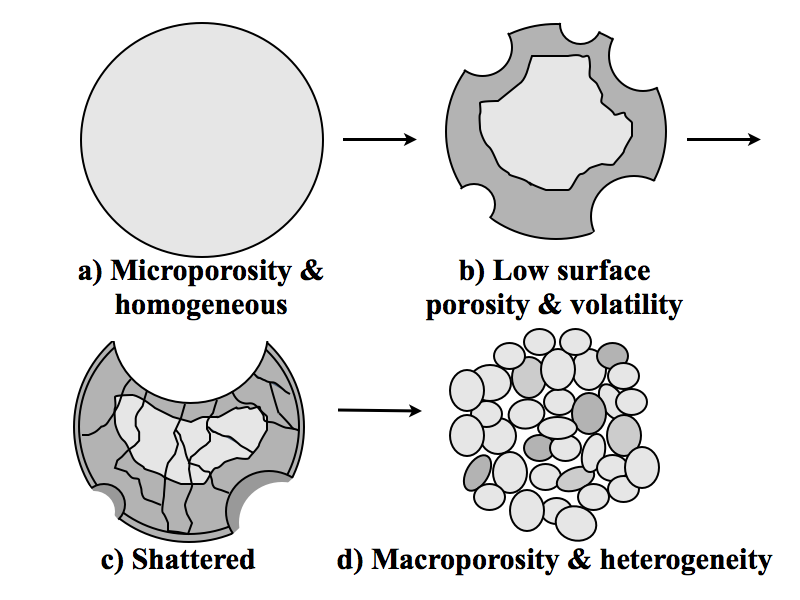}
 \caption{\small Schematic showing a possible evolutionary track for a
   KBO. a) A cross section through a primordial planetesimal with low
   bulk density and high microporosity. b) After many small impacts,
   the bulk density has increased and volatile composition decreased
   at the surface. c) After a large sub-catastrophic collision, the
   body is shattered and the surface is covered with ejecta. d) A
   catastrophic impact event disrupts the body, creating a rubble pile
   with high macroporosity and heterogeneous internal composition.
   \label{fig:cartoon_evol}}
\end{figure}

%It is unclear whether all bodies in the Kuiper Belt including
%the largest ($\sim$ Pluto-sized) have been involved in significant
%collisional processing \citep[the potential family around EL 61, {\it
%  Brown} this volume, suggests large KBOs may have been involved in at
%least one collision but the dynamical simulations of ][suggest only
%objects less than 100 km have been envolved in catastrophic
%collisions]{Davis97}.  In any case 
 
The observations summarized in the previous section indicate that
collisions are an important factor in the evolution of the Kuiper
Belt.  In studying the physical effects of collisions between KBOs, we
are guided by the mature studies of collisions in the asteroid belt
\citep{Asphaug02,Holsapple02}. However, the possible outcomes of
collisions between KBOs are more diverse compared to asteroids because
of the dynamical state of the system and the range of physical
properties of individual KBOs.

The important observations that inform the range of possible collision
outcomes are as follows.  The dynamical state of the Kuiper Belt has
changed dramatically with time ({\em Morbidelli et al.}, this volume);
hence the mutual collision velocities between KBOs also varied with
time. In the modern Kuiper Belt, the mutual collision velocities are
around 1~km~s$^{-1}$ \citep{Trujillo01} for classical KBOs and
slightly higher for the other populations ({\em Gladman et al.} this
volume). For these impact velocities, most bodies smaller than 100's
km size have experienced a catastrophic disruption event, while most
of the larger bodies have survived. All bodies should have suffered
the production of a significant density of surface impact craters.
%(DIDN'T STERN
%PREDICT A SURFACE DENSITY OF CRATERS ON PLUTO?).

In addition to the dynamical impact conditions, the physical
properties of KBOs are important. While the present observations are
limited, the range of bulk densities of KBOs is $<$1 to $\sim$ 2.6 g~cm$^{-3}$ ({\em Stansberry et al.} this volume)
and the largest KBOs have a range of surface volatile compositions
(H$_2$O, CH$_4$, etc) in addition to a refractory (rock and organic) component
({\em Barucci et al.} this volume). From these observations and studies of short-period comets,
believed to be fragments from KBOs, a typical KBO has a significant
(but unknown) fraction of volatiles and high porosity. 
%{\bf FIX THIS
  %PARAGRAPH -- summarize what else is known about material
  %properties.}

Given the possible range of material properties and impact conditions,
we outline the potential array of dramatically different outcomes from
collisions between KBOs:

\begin{enumerate}
\item {\bf Surface Impact Features}: As a result of the rich
  collisional history of the Kuiper Belt, impact craters are expected
  to be common on the surfaces of KBOs. However, the morphologies and
  size distributions are sensitive to the surface and internal
  structure of the body. Some highly porous bodies survive the
  formation of multiple, large craters comparable to the radius of the
  object, as on the low-density asteroid Mathilde \citep{Veverka97}.
  Alternatively, only small craters may be observed on rubble piles
  formed by catastrophic disruption, such as asteroid Itokawa
  \citep{Fujiwara06}.
\item {\bf Surface Composition and Color}: Impact craters and
  catastrophic disruption events may darken the surface by removing
  volatiles via heating from the energy of the impact or brighten the
  surface of a body by excavating fresh ices (Fig. \ref{fig:color}).
\item {\bf Density of Surface Materials}: Laboratory impact craters in
  highly porous and compressible materials compact the impact site,
  creating density heterogeneity on the surface \citep{Housen99}.
  Over time, cumulative small impacts on a microporous surface may
  increase the bulk density and decrease the bulk porosity (Fig.
  \ref{fig:cartoon_evol}ab).  In contrast, modeling results indicate
  that nearly all ejecta from a crater in a macroporous body may reach
  escape velocities, leaving the bulk density unchanged
  \citep{Asphaug98}.
\item {\bf Internal Structure and Composition}: A sub-catastrophic
  impact may shatter a body (and create a large crater) but leave the
  original internal material relationships intact (Fig.~\ref{fig:cartoon_evol}c), while a
  catastrophic impact both shatters and disperses a body such that the
  gravitationally reaccumulated remnants are rubble piles with high
  macroporosity and mixed composition (Fig.  \ref{fig:cartoon_evol}d).
%  {\bf should we amend the figure to show shattered objects as well -
   % I'm thinking of Stickney and Asphaug's Ast III chapter}
\end{enumerate}

In the next section, we describe the factors that control the outcome
from individual collision events.

\section{FACTORS THAT CONTROL OUTCOMES OF COLLISIONS}\label{sec:fact}

Recent advances in the understanding of the physics of collisional
processes between icy, porous bodies provide new fuel to the study of
the role of collisions in the Kuiper Belt.  The outcome of collision
events are governed by the impact conditions (velocity, angle, and
mass of each body) and the physical properties of the colliding bodies
(strength, composition, and internal structure).  Both the impact
conditions and physical properties affect the efficiency with which
the energy of the impact is coupled to the target. In this section, we
summarize four overarching factors that control the outcome of an
impact event between KBOs. In the following section (\S
\ref{sec:studies}), we will describe laboratory and numerical
experiments on KBO analog materials that investigate these controlling
factors.

First, the composition and internal structure of the bodies determines the
critical velocity required to enter the strong shock regime, where the
deformation and coupling of energy and momentum can be described
through the Rankine-Hugoniot conservations equations (\S
\ref{sec:speed}). Slower impact events, where plastic deformation
dominates, require more detailed knowledge of the physical properties
(particularly strength) of the bodies compared to the strong shock
regime. Collisions between KBOs are likely to span the range of
plastic and shock deformation.

Second, the final outcome is a balance between the forces of strength
and gravity (\S \ref{sec:scaling}). Scaling laws have been developed
for cratering and catastrophic disruption in each regime, but a large
transition region exists. Because of the low gravity and expected low
strength of KBOs, many collisions may fall in the transition region.

Third, the internal structure and composition of the colliding bodies
may significantly affect collision outcomes (\S \ref{sec:porosity}).
Some of the impact energy will be partitioned into phase changes when
highly volatile materials are present.  High levels of porosity also
alter the energy coupling by acting as a shock absorber and
localizing shock deformation.  The momentum coupling with high
porosity changes the excavation flow in the cratering regime and the
dispersal of fragments in the disruption regime compared to collisions
between solid bodies.

Fourth, collision outcomes are sensitive to the mass ratio of the
projectile and target. At the same kinetic energy, a larger projectile is more efficient at
removing mass than a smaller
projectile (\S \ref{sec:mratio}).

\subsection{Shock Deformation}\label{sec:speed}

%In order to have a detailed discussion about sub and supersonic
%impacts we will first present a brief introduction to shock physics.

%{\bf tie this sectiont to: deformation (compaction, fragmentation),
 % devolatilization, and the primary factor leading to cratering or
%  shattering and disruption.}

%{\bf one of the subsections needs a basic Q*D = Q*S + grav binding
 % energy - see the AST III presentation of this - then you can talk
 % about gravity transition and mass ratio more easily}

In most high energy impact events, the deformation is driven by a
shock wave.  The energy from the shock controls the physical
deformation from the collision, such as fragmentation, pore collapse,
heating, and phase changes.  The shock also determines the deposition
of momentum that leads to crater excavation or dispersal of fragments
following a catastrophic disruption event. The amount of deformation
can be estimated by considering the volume of material shocked to a
given peak pressure. 
%{\bf could ref fig in Melosh book here}.

A strong shock wave is produced in a {\it hypervelocity} impact event,
where the impact velocity exceeds the bulk sound speed ($c_b$) in both
the target and projectile.  However, mutual collision velocities
between KBOs are likely to span the range from subsonic to supersonic
(hypervelocity) collisions (see below).  When collisions are
comparable to the sound speed, plastic deformation dominates, rather
than strong shock deformation.  Under subsonic conditions, collisions
are simply elastic and governed by the coefficient of restitution.

In this section, we present a summary of the shock physics that
determines the outcome in the hypervelocity regime. The peak shock
pressure is deduced from the conservation equations and material
equation of state, describing the pressure-volume-temperature
($P-V-T$) states. A shock wave satisfies the Rankine-Hugoniot (R-H)
mass, momentum, and energy conservation equations \citep{Rice58}:
\begin{eqnarray} \label{eq:rh}
  u_{i}-u_{0} & = & U_S \left( 1- \frac{V_{i}}{V_{0}} \right),
  \label{eq:mass} \\
  P_{i}-P_0       & = & \frac{U_S}{V_0} \left( u_{i} - u_{0}
  \right), \label {eq:mom} \\
  E_{i}-E_0 & = & \frac{1}{2} \left( P_{i} - P_0 \right) \left( V_0 - V_{i} \right). \label{eq:energy} 
\end{eqnarray} 
In the above formulae $u$ is particle velocity, $U_S$ is shock velocity, $V$ is specific
volume ($=1/\rho$, where $\rho$ is density), $P$ is pressure, and $E$
is specific internal energy.  The initial unshocked state is
subscripted $_0$ and the final shocked state is subscripted $_i$.
%It follows directly from the
%R-H equations that the energy of the shock wave is equally partitioned
%into internal energy and kinetic energy.

The shock Hugoniot is the curve that describes the locus of possible
$P-V$ shock states for a given initial $P-V-T$ state. For a given
impact scenario, the shock pressure is calculated using
Eq.~\ref{eq:mom}, the impact velocity, and the equations of state of
the target and projectile. Many materials may be described using a
simple linear $U_S-u$ shock equation of state of the form
\citep{Ruoff67},
\begin{equation}
U_S = c + s u, \label{eq:shockeos}
\end{equation}
where $c$ and $s$ are material constants \citep[for their relationship
to finite strain theory, see][]{Jeanloz89}. The linear shock equation
of state is simply a representation of the $P-V$ shock Hugoniot
translated into $U_S-u$ space using the R-H equations.

In the planar impact approximation \citep[also called the impedance
match solution, see derivation in][]{Melosh89}, the particle
velocities induced by the shock wave reduces the projectile's velocity
and mobilizes the target such that continuity at the projectile-target
interface is achieved and
\begin{equation}
u_{t} = v - u_{p}, \label{eq:vu}
\end{equation}
where $v$ is the impact velocity and subscripts $_t$ and $_p$ refer to
the target and projectile, respectively. The shock pressure is derived
by solving for $u_t$ using the equality of Eq.~\ref{eq:mom} in the
target and projectile and substituting Eqs.~\ref{eq:shockeos} and \ref{eq:vu},
\begin{equation}
\rho_{0,t}(c_t+s_t u_t)u_t = \rho_{0,p}(c_p+s_p (v-u_t))(v-u_t).
\end{equation}
The quadratic function for $u_t$ is readily solved. In the case of
identical shock equations of state in the target and projectile, the
particle velocity is equal to $v/2$, and the peak pressure is given by
$\rho_0(c+sv/2)(v/2)$. Because of strength and phase changes, the
$U_S-u$ shock equations of state for natural materials are usually fit
with multiple linear segments in $u$, corresponding to different
pressure ranges on the shock Hugoniot. The shock equations of state
for many rocks and minerals are compiled in
\citet{ahrens95a,ahrens95b}, and the equations for nonporous and
porous H$_2$O ice are given in \citet{Stewart04,Stewart05}.

As the shock wave propagates into the target, the peak pressure,
derived from the planar impact approximation, decays from rarefaction
waves on the free surfaces. The size of the region at peak pressure
(known as the isobaric core) and the decay exponent depend on the
impact velocity and material properties (e.g., equation of state and
porosity) \citep{ahrens87,Pierazzo97}.  In general, the pressure decay
is steeper for high velocities because of energy partitioning into
phase changes.  The occurrence of impact-induced phase changes can be
estimated by considering the critical shock pressures required for
melting and vaporization. When the shock pressure is above a critical
value, the material is melted/vaporized after passage of the shock
wave and return to ambient pressure conditions.

The present mean mutual collision velocity between classical KBOs
($\sim$1~km~s$^{-1}$) is lower than the bulk sound speed of full
density silicates and ices.  Nonporous H$_2$O ice has a $c_b$ of 3.0
km~s$^{-1}$ at 100~K \citep{Petrenko99,Stewart05}.  Silicate rocks
have larger $c_b$, typically around 5 km~s$^{-1}$ \citep{Poirier00}.
Sounds speeds of laboratory preparations of nonporous ice-silicate
mixtures, with up to 30\% by weight sand, are similar to pure H$_2$O
ice \citep{Lange83}.  Pure porous H$_2$O ice, on the other hand, can
have much lower sound speeds, from 0.1 to 1.0 km~s$^{-1}$ for bulk
densities of 0.2 to 0.5 g~cm$^{-3}$ \citep{Mellor75,Furnish97}.
Silica aerogels with densities of about 0.2 g~cm$^{-3}$ have sounds
speeds of about 200 m~s$^{-1}$ \citep{Gross88}, and 35\% porous sand
has a sound speed of 130 m~s$^{-1}$. If KBOs are volatile rich and
porous, then mean present-day collisions may be supersonic. During the
collisional evolution of the Kuiper Belt, collisions span the subsonic
to supersonic regimes.

Understanding the controlling physics in the subsonic regime, where
plastic deformation dominates, will require focused studies on analogs
for the range of mechanical structures in the Kuiper Belt (\S
\ref{sec:studies}).  In the strong shock regime, crater scaling
relationships and catastrophic disruption theory are applicable, as
described in the next section.

\subsection{Strength and Gravity}\label{sec:scaling}

%{\bf tie to surface features (craters) and internal structure, surface
 % composition and color}

The final outcome of a collision, e.g., crater size or dispersed mass,
depends on the balance between strength and gravitational forces. In
the case of impact cratering, the relationship between the size and
velocity distribution of the impacting population and the observed
crater population can provide insight into the collisional history of
a system (as has been done for the terrestrial planets \citep{Strom05}
and asteroids \citep{Obrien05}). Backing out the impactor properties
requires the application of the appropriate crater scaling
relationships, which depend on both the impact conditions and material
properties. In the case of catastrophic disruption, knowledge of the
properties of the populations of disrupted and primordial bodies
provide strong constraints on the collisional evolution of the Kuiper
Belt. Here, we present the crater size and catastrophic disruption
scaling laws in the strength and gravity regimes.

\subsubsection{Crater Scaling Theory}

\begin{figure}[h]
  \includegraphics[width=80mm]{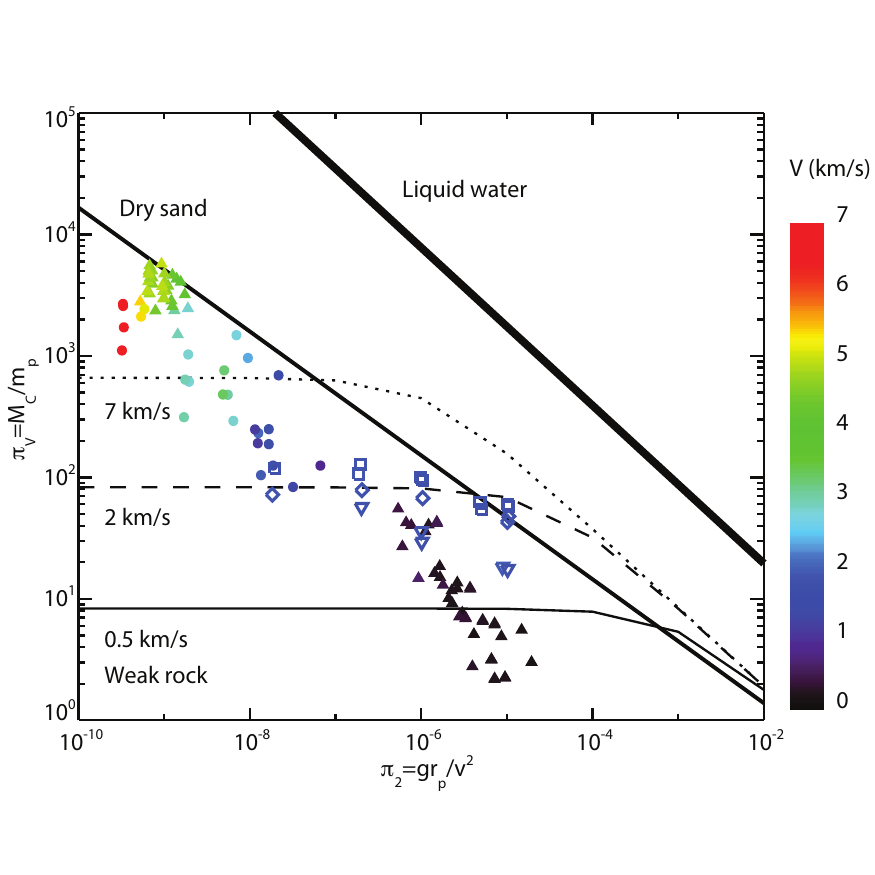} 
 \caption{\small Cratering efficiencies (ratio of ejected and
   displaced mass to projectile mass, $\pi_V$) as a function of the
   ratio of gravitational to inertial forces (inverse Froude number,
   $\pi_2$) for different target materials at different velocities.
   Lines denote fitted cratering efficiencies in liquid water, dry
   sand (35\% porosity), and weak rock using Eq.~\ref{eq:piscale}.
   Experimental data (see text): nonporous ice $\blacktriangle$; 50\%
   porous ice $\bullet$; various crushable non-icy mixtures with
   porosities of about 40\% $\square$, 70\% $\Diamond$, and 96\%
   $\bigtriangledown$.  Colors denote impact velocity.
   \label{fig:piscale}}
\end{figure}

Development and validation of the appropriate scaling relationships is
crucial for the next generation of collisional evolution models of
KBOs that include consideration of physical deformation effects.
Because of their low gravity and likely low strength, the outcome of
collisions between KBOs is near the transition between the strength
and gravity regimes.  In this section, we discuss the strength to
gravity transition and summarize the equations and material parameters
for crater scaling for comparison to laboratory craters in ice and
porous targets in \S \ref{sec:crater}.

In a cratering event, the shock-driven excavation flow produces a
roughly hemispherical cavity, called the transient crater.  Assuming
that material strength can be represented by a single parameter, $Y$,
the transition size between the strength and gravity-controlled
cratering regimes is proportional to $Y/\rho g$.  $Y$ is the dominant
strength measure that controls crater size (e.g., shear strength);
$\rho$ is the bulk density of the target; and $g$ is the gravity of
the target \citep{Melosh77,Melosh78}. As the impact energy increases,
the outcome of collisions will transition from a cratering regime to a
total body disruption regime. The criteria for catastrophic
disruption, $Q^*_D$, is defined as the specific energy (kinetic energy
of the projectile divided by the mass of the target) required to
disrupt and gravitationally disperse half the mass of the target
\citep{Melosh97}.  Note that, unlike the cratering regime, disruption
is governed by the bulk tensile strength of the body, which is
typically an order of magnitude lower than the compressive strength of
brittle solids (see \S \ref{sec:CDTheory} \& \ref{sec:mratio}).

The theory for crater size scaling based on impact parameters and
material properties is summarized by \citet{Holsapple93}. A common
approach utilizes $\pi$-scaling, with empirical constants derived from
impact and explosion cratering experiments under Earth's gravity and
high gravity. Predicting the final crater volume and shape requires
two steps: (i) calculating the volume of the transient crater cavity
using the $\pi$-scaling laws and (ii) calculating the amount of
collapse of the transient crater to the final crater volume and shape.
The first step is better understood than the second.

In $\pi$-scaling, the cratering efficiency, $\pi_V$, is defined as the
ratio of the mass of material ejected and displaced from the transient
crater cavity to the mass of the projectile:
\begin{equation}
  \pi_V = \frac{\rho V}{m_p} = \frac{M_c}{m_p}, \\
\end{equation}
where $V$ is the volume of displaced and ejected material, $m_p$ is
the mass of the projectile, and $M_c=\rho V$. For strength-dominated
craters, the cratering efficiency depends on the ratio of a measure of
the shear strength of the target, $\bar Y$, to the initial dynamic
pressure, given by
\begin{equation}
  \pi_{\bar Y} = \frac{\bar Y}{\rho v^2_{\perp}},
\end{equation}
where, $v_{\perp}=v \sin \theta$, $v$ is the impact velocity, and
$\theta$ is the impact angle from the horizontal.  In the
gravity-dominated regime, the cratering efficiency depends on the ratio
of the lithostatic pressure at a characteristic depth of one
projectile radii, $r_p$, to the normal component of the initial
dynamic pressure (the inverse Froude number):
\begin{equation}
  \pi_{2}= \frac{g r_p}{v_{\perp}^2},
\end{equation}
where $g$ is the gravitational acceleration.

Impact experiments demonstrate that the transition from strength to
gravity-dominated cratering spans about two decades in $\pi_2$.
Following \citet{Holsapple93} and \citet{Holsapple04}, the cratering
efficiency can be defined by an empirical, smoothed function of the
form
\begin{equation}
  \pi_{V}= K \left( \pi_2 + \pi_{\bar Y}^{\beta / \alpha}
  \right)^{-\alpha}, \label{eq:piscale}
\end{equation}
where the exponents are related to a single coupling exponent, $\mu$,
by $\alpha=3\mu / (2+\mu)$ and $\beta = 3\mu/2$. The coupling exponent
$\mu$ is bounded by two cratering regimes: momentum scaling (where
$\mu=1/3$) and energy scaling ($\mu=2/3$)
\citep{Holsapple87a,Holsapple87b}. Note that Eq.~\ref{eq:piscale}
assumes that the target and projectile have the same density.  The
transition from strength to gravity dominated regimes occurs when
$\bar Y \sim \rho g r_p$.

In Fig.~\ref{fig:piscale}, cratering efficiencies are presented for
liquid water (K=0.98, $\mu=0.55$, $\bar Y=0$ MPa), dry sand (K=0.132,
$\mu=0.41$, $\bar Y=0$ MPa, 35\% porosity), and weak rocks (K=0.095,
$\mu=0.55$, $\bar Y=3$ MPa) \citep[values from][]{Holsapple04}. Dry
sand is a non-crushable porous material and weak rock is a reasonable
analog for nonporous H$_2$O ice. Cratering efficiencies in crushable,
porous materials, from hypervelocity experiments in vacuum, lie a
factor of few below the dry sand line \citep{Schultz05}. The
transition from strength regime (lower values of $\pi_2$, when
$\pi_{\bar Y}>\pi_2$) to gravity regime (higher values of $\pi_2$)
corresponds to the transition from a horizontal line in a
$\pi_V-\pi_2$ plot, when the cratering efficiency is independent of
$\pi_2$, to a power law with slope $-\alpha$. The cratering efficiency
in the strength regime increases with increasing impact velocities, as
indicated by the curves for impacts into weak rock targets at 0.5,
2.0, and 7.0 km~s$^{-1}$.  Note that the cratering efficiency in dry
sand is less than for weak rock in the gravity regime because of
energy dissipation in the porous sand. Data from impact cratering
experiments conducted in vacuum under Earth's gravity into nonporous
ice \citep[$\blacktriangle$,][]{Cintala85,Lange87,Burchell05} and 50\%
porous ice \citep[$\bullet$,][]{Koschny01,Burchell02} fall in the
strength regime. Cratering exeriments at 1.86~km~s$^{-1}$ in porous
mixtures of sand and perlite bonded with fly ash and water under one
atmosphere of pressure and varying gravity are nearly independent of
$\pi_2$, indicating strength-dominated behavior with a plausible
intersection with the gravity-dominated regime (the dry sand line)
\citep[\S~\ref{sec:compcr},][]{Housen03}.

Cratering events in the Kuiper Belt by a nominal 0.5~m radius body at
1 km~$s^{-1}$ onto targets of 0.1 to 1000 km radii correspond to
$\pi_2$ values in the range from $10^{-10}$ to $10^{-6}$. Therefore,
for the range of impact velocities in the Kuiper Belt,
Fig.~\ref{fig:piscale} demonstrates that the presence of any strength
is likely to control the final crater size for the majority of impact
events. In the upper range of possible values of $\pi_2$, gravity may
dominate if cratering is less efficient in KBOs than in dry sand.

After formation of the transient crater cavity by the excavation flow,
most craters undergo collapse to a final crater shape
\citep{Melosh99}.  For simple, strength-dominated craters, the final
crater size is similar to the transient cavity with some collapse of
the crater walls. For cratering in soils and rocks, the rim radius of
the transient crater is approximately $R_r=1.73 V^{1/3}$
\citep{Holsapple93}. Complex, gravity-dominated craters undergo
significant collapse of the transient cavity, and the final crater rim
radius scales with the transient crater rim radius and gravity by
$R_{\rm complex}({\rm cm})=0.37R_r({\rm cm})^{1.086}(g/g_{\rm
  Earth})^{0.086}$ \citep{Holsapple93}. 

The final state of crater formation in porous, icy bodies is not well
understood. In \S \ref{sec:studies}, we discuss some of the laboratory
experiments that provide our best guesses at the appropriate crater
size scaling laws for the Kuiper Belt.

\subsubsection{Catastrophic Disruption Theory}\label{sec:CDTheory}

\begin{figure}
 \includegraphics[width=80mm]{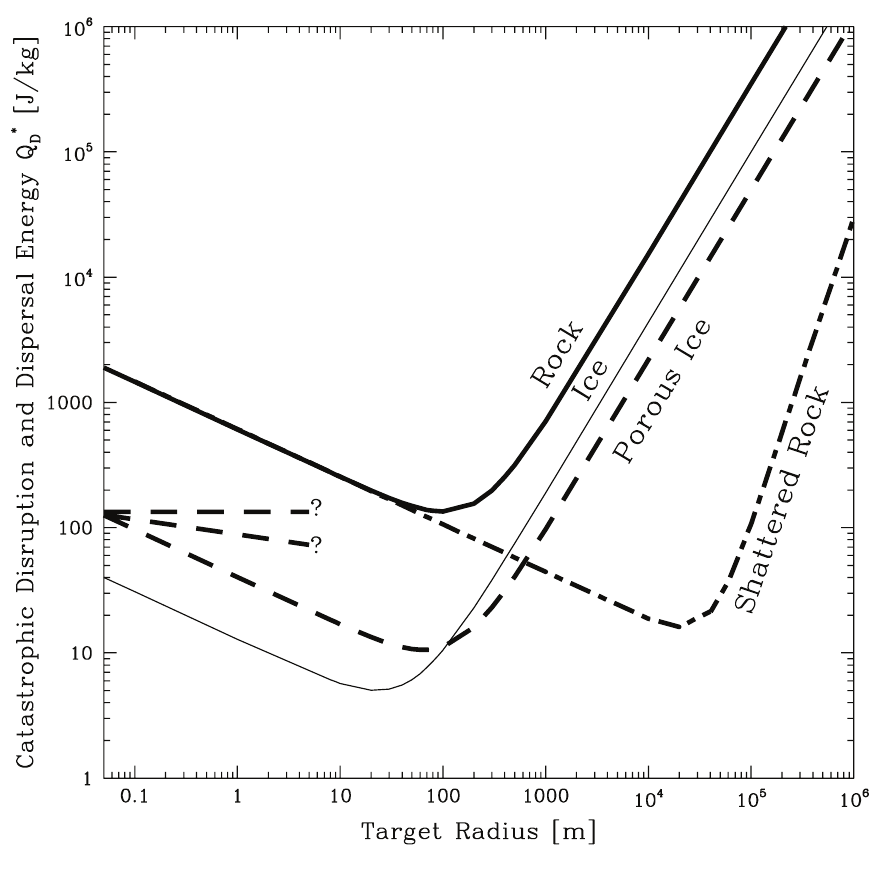} 
 \caption{\small Catastrophic disruption and dispersal energy
   (Q$_D^*$) versus target radius for rock (thick solid line),
   nonporous ice (thin solid line), 50\% porous ice (dashed line).
   The negatively sloped portions of the curves are in the strength
   regime, the positive slopes are in the gravity regime. The criteria
   for rock is based on angle averaged results from 3 km s$^{-1}$
   collisions onto basalt \citep{Benz99}. The nonporous and porous ice
   intercepts are based on laboratory experiments \citep[][Q$_D^*$ =
   40 J kg$^{-1}$ and Q$_D^*$ = 143 J kg$^{-1}$ for low and high
   porosity targets, respectively, of 5 cm radius]{Arakawa02}.  The
   extrapolation into the gravity regime is highly uncertain for
   porous materials. These results assume energy coupling by a small
   projectiles compared to the size of the target.
   \label{fig:qstar}}
\end{figure}

Based on the models of the collisional evolution of the Kuiper Belt,
it is probable that a large fraction of bodies have suffered both
cratering events and disruptive collisions.  The catastrophic
disruption criteria, $Q_D^*$, is the ratio of the projectile's kinetic
energy to the mass of the target required to disrupt and disperse half
the mass of the target. The criteria has two components \citep{Davis79}, 
\begin{equation}
Q_D^* = Q_S^*+Q_b,
\end{equation}
where $Q_S^*$ is the strength of the body to shattering and $Q_b$ is
the gravitational binding energy of the target. 

The catastrophic disruption criteria for a rocky body is shown by the
thick solid line in Fig.~\ref{fig:qstar} \citep[\S\ref{sec:CDnum} and
][]{Benz99}. The critical energy is averaged over all impact angles. A
head-on collision is most efficient, requiring about an order of
magnitude less energy compared to the angle average
\citep{Benz99,Leinhardt00,Leinhardt02}. In the strength regime, where
$Q_S^*$ dominates, the critical energy decreases with increasing
target size because tensile strength, the controlling strength
measure, is scale dependent \citep{Housen99b}. The larger the body,
the larger the number of pre-existing natural flaws and the lower the
tensile strength. In the gravity regime, pressure from the self
gravity of the object increases the strength, following the shattered
rock curve for head-on impacts \citep{Melosh97}.  In the gravity
regime, the gravitational dispersal criteria dominates over shattering
by orders of magnitude.  Note that the standard disruption criteria
curves assume that the size of the projectile is small compared to the
target (see \S\ref{sec:mratio}).

The manner in which volatile content and porosity affect the
disruption criteria is not well understood. Here, we estimate the
effects of each using nonporous and porous H$_2$O ice as an example.
There has been little work on catastrophic disruption of large objects
in the gravity regime at impact speeds and compositions that are
relevant to the the present day Kuiper Belt, thus this discussion is
meant to provide general guidance not detailed values.

In the strength regime, the $Q_D^*$ intercepts for nonporous and
porous ice (thin solid line and dashed line, respectively) are tied to
results from laboratory disruption experiments \citep[see \S
\ref{sec:CDlab} and][]{Arakawa02}.  These values are consistent with
other experimental results \citep{Ryan99}. The slope of $Q_S^*$ for
pure ice is assumed to be the same as for rock. For porous ice, on the
other hand, the slope in the strength regime is particularly
uncertain. The slope for porous materials may be much shallower than
for nonporous materials because the size-dependent scaling of flaws may not
apply \citep{Housen99b}. This uncertainty is depicted in
Fig.~\ref{fig:qstar} by several dashed lines of varying slope.
Perhaps counterintuitively, in the strength regime, a porous material
is harder to disrupt than a nonporous material due to localization of
energy by compaction of pores and/or reflection of the shock wave off
of free surfaces (see \S \ref{sec:CDlab}).

In the gravity regime, the $Q_D^*$ criteria for nonporous ice lies
below the rock criteria by the ratio in mass (for this plot vs.\ 
target size).  This is consistent with the numerical impact
simulations into ice targets by \citet{Benz99}.  Similarly, a porous
target of the same size is easier to disrupt because of its lower
total mass. Adjusting the gravity-dominated $Q_D^*$ criteria by the
ratio in total mass makes the unreliable assumption that the energy
coupling from the collision is similar for each material. Because of
the significant dissipative effects of porosity, porosity may have a
large affect on energy coupling in the gravity regime (the $Q_S^*$
term may be more significant). More work is needed to determine
exactly how porosity affects the energy coupling for catastrophic
disruption events.

As with impact cratering events, it is clear that in order to predict
the collision outcome from a disruption event, it is necessary to know
something about the material properties of the KBOs. In the case of
catastrophic disruption global properties, rather than surface
properties, are more important: for example, is the target porous,
icy, rocky? In \S \ref{sec:cat}, we discuss the small amount of work
on catastrophic disruption of KBO analog materials.

\subsection{Internal Structure and Composition}\label{sec:porosity}

KBOs are likely to possess a wide variety of internal structures, as
depicted in Fig.~\ref{fig:cartoon_evol}.  Dynamical excitation and
increased collision frequencies from the migration of Neptune removed
most of the mass from the original Kuiper Belt, leaving a mixture of
collision fragments and unmodified material \citep[][{\em Morbidelli
  et al.} this volume]{Davis97,Hahn99}.  Comets may provide clues to
the present internal structure of KBOs; however, comets are expected
to be diverse themselves \citep[{\it Barucci et al.} this
volume,][]{Weissman05}.

Porosity, either primordial or the result of collision events, is a
major complicating factor in predicting the amount of shock
deformation.  Since KBOs are expected to contain a range of
porosities, the outcome of individual collisions could vary widely
depending on the properties of the colliding bodies. When the initial
porosity is high, the shock impedance (the bulk density times the
sound speed) is low, and the peak shock pressures produced for a given
impact condition are lower compared to a solid target (Equations
\ref{eq:rh} \& \ref{eq:mom}). For a given shock pressure, however, the
internal energy increase is larger in a porous material because of the
greater change in volume during shock compaction (Equation
\ref{eq:energy}).  Hence, the temperature rise due to a shock is
higher in porous materials, and impacts into porous ices may result in
abundant melting or vaporization near the impact site
\citep{Stewart05,Stewart04}.  Porosity is an efficient dissipator of
shock energy.  As the shock propagates into the target, porosity
increases the decay rate of the shock because of the increased energy
partitioning into heat \citep[e.g.,][]{Meyers01}.  Therefore, the
shock-deformed volume in porous materials is smaller compared to a
solid.

The length scale of the porosity is also important. Small-scale
porosity compared to the shock thickness is described as {\it
  microporosity}. The thickness of the shock wave is proportional to
the scale of the topography on the surface of the projectile.
Large-scale porosity, e.g. a rubble pile of solid pieces, is described
as {\it macroporosity}. In the latter case, the solid (e.g.,
monolithic) pieces may have high strength, and impact cratering events
onto a monolithic piece would reflect the high surface strength. For
catastrophic disruption events, however, a rubble pile has low bulk
tensile strength.  In a rubble pile, the shock wave would reflect upon
encountering void space between solid boulders, and as a result, the
energy from the shock would be deposited in a smaller volume compared
to a shock wave propagating through a monolith of competent rock. On
the other hand, a microporous body may have low surface compressive
strength, but because of the efficient shock dissipation, a more
energetic impact is required to catastrophically disrupt the body
\citep{Asphaug98}. Hence, both macroporous and microporous bodies may
have high {\it disruption strength}.

Compositional variation and surface layers also change the way energy
is coupled into the target. The impact energy will be partitioned into
more compressible phases and a larger $Q^*_D$ is required to disrupt a
more compressible material \citep{Benz99}.  Because some of the energy
of the impact is partitioned into heating, each collision event will
also result in net devolatilization.  Finally, phase changes (melting,
vaporization) of volatile materials will result in steeper decay of
the shock wave \citep{Ahrens77,Pierazzo97,Pierazzo00} that tends to
localize the shock deformation in a manner similar to the effects of
porosity.

\subsection{Mass Ratio}\label{sec:mratio}

\begin{figure}
 \includegraphics[width=80mm]{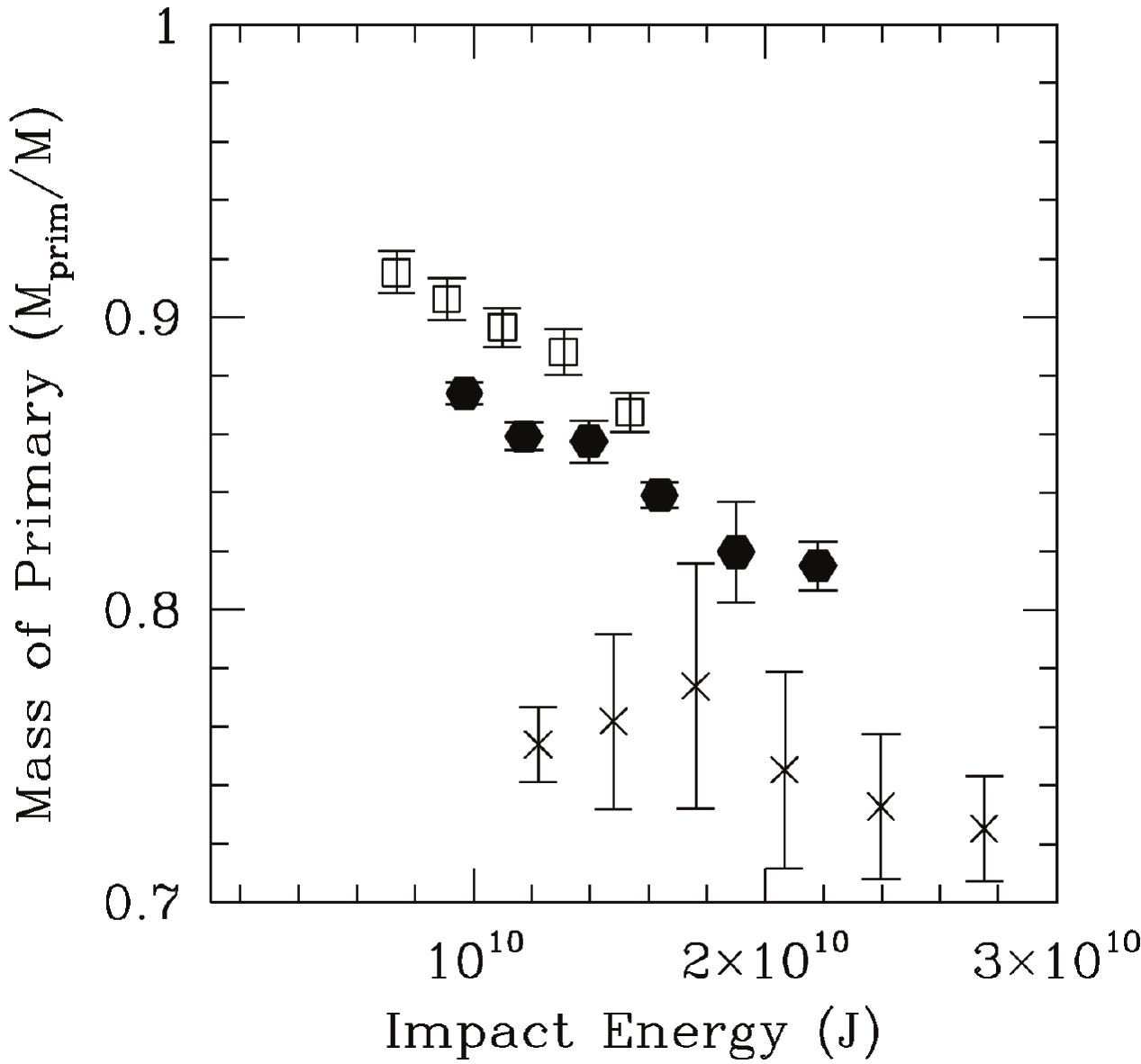} 
 \caption{\small Mass of the largest reaccumulated body ($M_{prim}$ in units of total system mass $M= M_{proj} + M_{targ}$) following a
   large collision event as a function of the kinetic energy of the
   projectile. Results from $N$-body impact simulations of km-sized
   rubble piles in the subsonic regime \citep{Leinhardt02}. The
   crosses, filled hexagons, and open squares denote projectile to
   target mass ratios of 1:3, 1:6, and 1:9, respectively. In all
   cases, targets were identical with a radius of 1 km. All data
   points are averaged over several simulations at various impact
   angles (rms error bars). Note that, for the same kinetic energy,
   the largest projectiles produced the smallest largest
   post-collision remnant. These results indicate that increasing the
   projectile size increases the disruption efficiency.
   \label{fig:mratio}}
\end{figure}

The mass ratio of the colliding bodies is also an important factor in
determining the collision outcome.  In simulations of subsonic
collisions, \citet{Leinhardt02} found that smaller projectiles were
not as efficient at disrupting targets as larger projectiles
\citep[see Fig.~\ref{fig:mratio} from][]{Leinhardt02,Melosh97}. The mass ratio affects the volume over which the impact energy and momentum are deposited. When the projectile is much
smaller than the target, the impact directly affects a small volume,
about the size of the projectile. The rest of the target acts to
dampen any material motion.  When the projectile mass is similar to
the target mass, on the other hand, the projectile comes in direct
contact with a significant volume fraction of the target.  As a
result, the specific energy required to catastrophically disrupt a
target decreases by orders of magnitude.

The dependence on mass ratio has not been studied directly for
hypervelocity impacts, although the dependence on the coupling of
energy and momentum should be similar to the subsonic case.  First,
the size of the peak pressure region (isobaric core) is proportional
to the size of the projectile.  Second, the decay of the peak shock
pressure with distance depends on the impact velocity.  The decay is
steeper for higher impact velocities because more energy is
partitioned into phase changes and deformation. Low impact velocities
have a more shallow decay exponent (in the elastic limit). The
particle velocities are proportional to the peak shock pressure; thus
the shock pressure profile in the target will affect the dispersal of
fragments and the catastrophic disruption criteria. $Q_D^*$ should
decrease as the projectile size increases for a fixed kinetic energy
of the projectile.

Let us now consider a likely impact scenario in the Kuiper Belt.  For
an example 100 km volatile rich target in the Kuiper Belt,
Fig.~\ref{fig:qstar} predicts that about $10^5$ J kg$^{-1}$ is
necessary for catastrophic disruption. At 1 km s$^{-1}$, the
projectile would have a radius two thirds that of the target. However,
Fig.~\ref{fig:qstar} assumes small (point source) projectiles. As the
projectile to target mass ratio approaches unity, the amount of energy
per target mass needed to disrupt the target drops
\citep{Holsapple93,Melosh97,Leinhardt02}.  Therefore, a 100 km target
may indeed be catastrophically disrupted by a smaller projectile than
predicted in Fig.~\ref{fig:qstar}, and the larger objects in the
Kuiper Belt may have suffered more catastrophic or near catastrophic
impacts than inferred in previous studies. More work is needed to
determine how $Q_D^*$ behaves with mass ratio in hypervelocity
collisions.

\section{STUDIES OF COLLISIONS IN ANALOGS TO KUIPER BELT
  OBJECTS}\label{sec:studies}

We now turn to laboratory and numerical experiments on the major
factors that affect the collision outcomes described above.
Laboratory experiments and numerical models of collisions between icy
and/or porous bodies serve as the best analogs at present for impacts
into KBOs.  There is a large body of laboratory work on cratering
impacts into volatile ices and mixtures as well as porous material (\S
\ref{sec:crater}). Although a coherent theory will require additional
experiments, the laboratory results are a good guide to the possible
outcomes of cratering collisions on KBOs.  In comparison, the
laboratory and numerical experiments on catastrophic disruption of KBO
analogs are more limited (\S \ref{sec:cat}). This is due in part to the
inability to study catastrophic disruption in the gravity regime in
the laboratory and the difficulties in modeling collisions between
porous, volatile bodies.

\subsection{Cratering}\label{sec:crater}

We begin with the results of several impact
cratering studies into nonporous H$_2$O ice and ice-silicate mixtures
(\S \ref{sec:nonp}).  Then, the effects of porosity are introduced (\S
\ref{sec:porous}).  However, it is difficult to deconvolve the effects
of porosity and low material strength in laboratory studies. Possible
outcomes include cratering events that result in compaction rather
than the normal crater excavation flow (\S \ref{sec:compcr}).
Relatively little work has been conducted on ices more volatile than
H$_2$O, which have been observed on the surfaces of the largest KBOs
({\em Barucci et al.} and {\em Brown} this volume) (\S \ref{sec:vol}).
Finally, experiments into targets with surface layers of different
strength materials can also have a significant affect on the crater
morphology (\S \ref{sec:layer}). Because of the influence of an
atmosphere on the final crater form, explosion cratering studies
\citep{Holsapple04} are not included in this discussion.

\subsubsection{Cratering in nonporous H$_2$O ice and ice-silicate mixtures}\label{sec:nonp}

As a low density and volatile material, solid H$_2$O ice represents a
very simple model for the bulk properties of KBOs. Depending on the
evolution of KBOs, some surfaces may be dominated by solid ice.  Most
laboratory impact experiments in ice are conducted in the strength
regime.  Generally, cratering efficiencies in solid ice are similar to
a dry soil or weak rock (Fig.~\ref{fig:piscale}) \citep{Chapman86}.
Solid ice cratering experiments span impact velocities of 0.1 to 7.3
km~s$^{-1}$ using a wide range of projectile materials
\citep{Burchell05,Croft79,Lange82,Lange87,Shrine02,Grey03,Kawakami83,
  Kato95,Iijima95,Cintala85}.  For a given impact energy, the crater
volume is more than an order of magnitude larger than craters formed
in a typical hard silicate rock.  In many of these experiments, the
measured volumes of the craters include a component of spalled
material (near-surface material ejected under tensile failure),
forming a terraced crater morphology with a central pit.  Hence, the
reported volumes are larger than the transient crater volume and
comparisons to $\pi$-scaling laws must be made with caution.  Also,
differences in ice target preparation contribute to scatter between
experiments.

Impact cratering experiments in solid H$_2$O ice have investigated the
effects of target temperature on the cratering efficiency
\citep{Lange82,Lange87,Grey03}.  Low-temperature ice has a cratering
efficiency between temperate ice and hard rock. Under fixed impact
conditions, the crater depth and volume decreased by factors of 2 and
4, respectively, as the ice temperature decreased from 253 K to 100 K
\citep{Grey03}.  It is well established that the yield strength of ice
increases as the temperature decreases \citep[e.g.,][]{Sammonds98};
however, the magnitude of the effect is not well predicted
\citep{Grey03}.

%%%CHECK: get a copy of Frisch 1992 book - looks like micrometeorite experiments
Thermodynamic analyses of shock wave experiments on solid H$_2$O ice
at 100~K indicate that peak shock pressures of 1.6 and 4.1 GPa are
required to produce incipient and complete shock-induced melting,
respectively \citep{Stewart05}.  For pure ice on ice impacts, these
pressures are achieved at impact velocities of about 1 and 2
km~s$^{-1}$. If the bulk shock impedance of solid ice is similar to
porous volatile-refractory mixtures, mass melting of solid ice within
KBOs is only expected at the upper end of the range of collision
velocities within the Kuiper Belt \citep[0.5 to 3
km~s$^{-1}$][]{delloro01}.

Nonporous mixtures of ice and silicates (e.g., ice-saturated sand)
have also been studied in the strength regime
\citep{Croft79,Koschny01b}. The cratering efficiency decreases with
increasing silicate content.

%{\bf now a summary of how the above applies to KBOs}

\subsubsection{Cratering in porous material}\label{sec:porous}
  
Several groups have conducted cratering experiments into porous
materials \citep{Schultz05,Schultz85, Schmidt80,
  Housen99,Housen03,Koschny01}. The porous targets include porous ice,
sand, Ottawa flint shot, pumice and vermiculite. Hypervelocity impact
experiments, conducted under vacuum, into pumice powder with
porosities between 35 and 50\% follow a single gravity-controlled
crater scaling that is slightly less efficient than dry sand
(Fig.~\ref{fig:piscale}) \citep{Schultz05}.  However, for lower
velocity impacts, in the strength regime or in the transition between
strength and gravity-dominated cratering, the effects of porosity can
be significant (\S \ref{sec:compcr}).

Results from cratering experiments into $\sim$50\% porous
\citep{Koschny01, Burchell02} and solid ice targets indicate that the
displaced and ejected mass scales linearly with impact energy. In
other words, the crater volume is proportionally larger by the
difference in target density.  However, it is unlikely that this
result can be extrapolated to events with much larger impact energies
because of the considerable effects of vaporization on the final
crater size \citep{Holsapple07, Schultz05}.

%% CHECK: get a copy of Livingston 1968 from Wolbach

Because of the large increase in internal energy associated with shock
compaction of porous H$_2$O ice, the critical pressures required for
shock-induced melting are lower compared to solid ice.  From shock
wave experiments, \citet{Stewart04} find that shock pressures of only
0.3 to 0.5 GPa initiate melting in 40-45\% porous ice, and complete
melting is reached by 2 GPa.  These pressures correspond to impact
velocities in the range of 1 to 2.5 km~s$^{-1}$ for collisions between
porous ice bodies.  Although porous ice has lower shock impedance than
solid ice, the increase in internal energy from pore compaction results
in similar critical impact velocities for shock-induced melting.  If
KBOs have shock impedances greater than pure porous H$_2$O ice, as
expected if they are ice-refractory mixtures, then mutual collisions
under the present dynamical environment will result in abundant
melting of H$_2$O ice.

Therefore, shock-induced melting in porous targets may produce crater
cavities lined with quenched melt (rapidly cooled liquid that
solidifies as a glass). Quenched melt lined craters have been observed
in laboratory impact experiments into 50\% porous H$_2$O ice (nylon
projectiles at 0.9 to 3.8 km~s$^{-1}$) \citep{Koschny01} and 5 to 60\%
porous soda lime glass (glass projectiles at 4.9 to 6.1 km~s$^{-1}$
\citep{Love93}. In some cases, all the impact-generated melt was
ejected from the crater. Hence, cratering events onto porous KBOs may
produce solid ice ejecta fragments.

The depth of penetration of the projectile plays a significant role in
the cratering efficiency in porous materials. In hypervelocity
impacts, the impact angle determines the outcome. For example, the
cratering efficiency in compressible porous perlite granules
($\rho=0.2$ g~cm$^{-3}$) increases as the impact angle decreases from
90$^{\circ}$ to 30$^{\circ}$ \citep{Schultz05}.  Vertical impacts into
porous materials penetrate deeply into the target, resembling a deeply
buried explosion. Low angle impacts, in contrast, reach a more optimal
shallow depth of burial to produce a larger crater. In the low
velocity regime, an impedance mismatch between the target and
projectile will also influence the depth of penetration. A dense
projectile may experience little deformation and penetrate deeply,
resulting in less efficient cratering compared to a projectile with
density that matches the target. Interpretation of the cratering
record on KBOs will need to include the role of impact angle and the
depth of penetration in the final crater size.

In the case where the projectile is more dense than the target (e.g.,
a solid rock meteoroid impacting a porous KBO), the impact conditions
may be supersonic for the target but subsonic for the projectile. In
this case, the projectile is not significantly disrupted by the impact
event. Laboratory experiments show that the penetration depth
increases as the density contrast between the projectile and target
increases \citep{Love93}. Intact or melted nylon ($\rho=1.14$
g~cm$^{-3}$) and copper ($\rho=8.92$ g~cm$^{-3}$) projectiles were
recovered after impacts at velocities up to 7 km~s$^{-1}$ into 50\%
porous H$_2$O ice \citep{Koschny01,Burchell02}. The experimental
results suggest that dense meteoroids may embed themselves into the
surfaces of KBOs and comets. In a pathalogical example, a population
of compacted, devolatilized projectiles might be found embedded in the
surfaces of very porous, volatile KBOs.

Reliable numerical models of crater formation in porous materials have
been hindered by the difficulty in modeling the shock compaction of
porous materials \citep[e.g.,][]{Hermann1969, Johnson1991}. Some
general results can be drawn from the relatively few simulations to
date: (i) a proxy model for porosity using layers of solid ice and
void and the Autodyne code \citep{Burchell05}, (ii) a P-alpha crush up
model for sand using the CTH code \citep{Housen00}, and (iii) a new
$\epsilon$-alpha compaction model using the iSALE code
\citep{Wunnemann06}. In summary, the transient crater diameters in
porous materials are smaller but the crater is deeper than those in
nonporous media.  The lower bulk density of the porous target allows
the projectile to penetrate more deeply.  The shock wave is attenuated more quickly in porous material because energy is partitioned
into crushing pores.  These numerical experiments show that porous
crushable objects are more resilient to large impact than nonporous
objects because the damage from the impact is much more localized.
With these more advanced models of porosity, future work can address
the volume of material that experiences deformation (fragmentation,
devolatilization) from impact events in the Kuiper Belt.

\subsubsection{Compaction Cratering}\label{sec:compcr}

%{\bf SAY SOMETHING ABOUT CONTROVERSY CITE ASPHAUG ASTEROIDS III (COHESION
%ATMOSPHERE) and also include the reference to a corroborating study
%referenced in AST III}

Observations of an unusual main belt asteroid, 253 Mathilde, have
incited several studies on the role of porosity on impact cratering.
Imaged during a flyby of the NEAR spacecraft, Mathilde has a low bulk
density ($\sim$ 1.3 g cm$^{-3}$) and exhibits four large impact
craters with diameters larger than the mean diameter of the asteroid
\citep{Veverka97}. The large craters have no visible ejecta blankets
or raised rims. As a result of their size, the craters are very close
to each other and yet seem to show no evidence of interaction. The
unique characteristics of Mathilde suggest that the internal structure
of this C-type asteroid is different from other classes of main belt
asteroids in a fundamental way.

\citet{Housen99} and \citet{Housen03} conducted a series of cratering
experiments into compressible, porous material in an attempt to
explain the craters on asteroid 253 Mathilde. The authors suggest that
high microporosity (40-60\%) and high compressibility lead to a
phenomena they termed compaction cratering.

In their studies, the projectile and impact velocity was held
constant, and the target porosity and gravity (using a centrifuge)
were varied. In the high gravity environment, the craters had no
raised rims and minimal ejecta outside the crater because most of the
ejecta never escaped the crater cavity. A computed tomography scan of
the crater showed a region of pore compaction approximately one crater
radius below the crater.  \citet{Housen99} and \citet{Housen03} also
impacted one of the used targets close to the original crater and
confirmed that there was little interaction between the craters. For
example, the first crater did not collapse as a result of the second
impact, nor was the first crater erased as a result of shaking or
ejecta filling in the first crater.

The authors conclude that large impacts onto compressible, highly
porous targets may not reach the gravity regime in which the gravity
scaling laws can be employed to predict crater diameter and depth. In
Fig.~\ref{fig:piscale}, the compaction craters in perlite and mixtures
of sand, perlite and fly ash (open symbols) are strength-dominated. As
a result, a porous, compressible object may have a very high
resistance to disruption even if both the tensile and compressive
strengths are low.

The occurrence of compaction cratering in nature is not understood and
presently a subject of debate. 
%These experiments were conducted under
%one atmosphere of pressure, which may have inhibited the excavation of
%the crater. 
More work on the compaction cratering phenoma is needed.
If compaction cratering is prevalent, the bulk density of a porous
compressible object may be significantly increased over the age of the
solar system by compaction from impacts.

\subsubsection{Cratering in other volatile materials}\label{sec:vol}

KBOs show wide diversity in volatile content. Large KBOs that are
bright enough for detailed spectroscopic study show evidence of
significant volatile content (for example, methane and ethane ices)
\citep[{\it Barucci et al.} this volume, ][]{Brown06,Barucci05}.
Laboratory experiments have found that the addition of material more
volatile than H$_2$O ice, such as CO$_2$ and NH$_3$ \citep{Burchell98,
  Burchell05,Schultz96} can increase the strength of the target and as a result decrease the cratering efficiency. 

The phase of the volatile is also important. Comet nuclei and their
precursors may contain trapped pockets of gas under high internal
pressures. If an impact event releases trapped gas, the vapor
expansion may aid in the ejection of more mass than would be possible
from the kinetic energy of the impact itself
\citep{Durda03,Schultz05,Holsapple07}.

\subsubsection{Cratering in layered targets}\label{sec:layer}

%{\bf Cratering in layered targets see morphological changes and if top
%  layer is thin, energy coupling effects(crater size) Quaide and
%  Oberbeck 1968; Gault et al 1968; Schultz 2000.}
  
\citet{Belton07} suggest that all three Jupiter Family Comet nuclei
(believed to originate in the scattered disk component of the Kuiper
Belt) that have been closely observed to date (Wild 2, Borrelly, and
Temple 1) show evidence of layering. \citet{Belton07} propose that
this layering is primordial and a result of the accretion process. By
extrapolation, the precursor objects in the Kuiper Belt may also have
layered structures.  Whether the observed layering is primordial or
not is a matter of debate; however, surface layering (a devolatilized
``crust'') was predicted for comets based on thermal evolution models
\citep{Belton99}.  Layering of different strength materials does
explain features seen on other objects in the solar system. For
example, concentric crater morphology on the moon and slightly filled
in linear structures on the asteroid Eros can be explained by regolith
overlying more competent rock.

\citet{Oberbeck67} and \citet{Ryan91} conducted experiments on
regolith covered targets and determined that the morphology of the
resulting crater changed depending on the depth of the regolith. This
result has been confirmed with numerical experiments by
\citet{Senft06}. \citet{Ryan91} conducted drop tests to study the
collision outcome of aggregate projectiles impacting different depths
of regolith (fine particles overlaying a concrete surface).  When the
depth of the regolith was at least the size of the projectile, the
aggregate lost $<$10\% of its mass when dropped from a height that
resulted in catastophic disruption when the surface was not covered
with a layer of fine particles. The porous regolith was very efficient
at dissipating the impact energy. 

In addition, impact experiments into granular mixtures of H$_2$O ice,
CO$_2$ ice, and pyrophylite that have experienced thermal
stratification produce craters with very different morphologies
\citep{Arakawa00}.  Finally, \citep{Schultz03} looked at the effect of 
layering on crater scaling. Craters retain their original diameter
until the layer becomes less than twice the projectile diameter (for
vertical impacts) or less than a projectile diameter (for oblique
impacts).  Even though the final crater depth is limited by the 
substrate, the diameter remains unaffected.  Imagery of craters on the
surfaces of KBOs would provide information about near-surface
layering.

%WHAT DOES
%THIS MEAN FOR KBOS? - TIE TO FUTURE OBSERVATION OF SURFACES and
%compositional maps?

%NEED TRANSITION FROM CRATERING TO CATASTROPHIC DISRUPTION - SHOULD BE SUMMARY OF CRATERING WORK

%There have been significant experimental work on cratering into
%volatiles, silicate mixtures, porous materials all of which could be
%analogs for KBOs. The morphology of the craters is affected in a
%complex manner by the material composition of the target. However, it
%is yet unclear if KBOs are dominated by the characteristics that have
%been highlighted in the impact experiments. More observational data on
%the surface composition of KBOs and impact experiments like Deep
%Impact would be invaluable in refining our current models for KBO
%composition.

\subsection{Catastrophic Disruption} \label{sec:cat}

%NEEDS AN INTRO - WILL TALK ABOUT BOTH LAB AND COMP EXPERIMENTS
%WHY IS THIS REGIME IMPORTANT TO THE KB?

As mentioned above, there has been much less work in the catastrophic
disruption regime than the cratering regime. We begin this section
with a brief summary of the laboratory experiments of catastrophic
disruption either using ice targets or investigating a range of
porosities. Next, numerical experiments on ice or porous
targets are presented.

\subsubsection{Catastrophic Disruption Laboratory Experiments} \label{sec:CDlab}

%ARAKAWA 2002 SHOULD BE ADDED TO THIS SECTION - RIGHT? YES.

Strength-regime laboratory experiments have investigated the
catastrophic disruption of icy and porous targets. Both \citet{Love93}
and \citet{Ryan99} conducted catastrophic disruption impact
experiments into macroporous targets. \citet{Arakawa02} performed
impact experiments into nonporous and porous pure ice and ice
silicate mixtures. In most of the experiments, the porous targets were
more difficult to disrupt because the kinetic energy of the projectile
is partitioned into crushing energy to fill void spaces \citep{Love93}
and the shock wave reflects off of the large number of free surfaces.
The result is significant attenuation of the shock wave compared to
solid materials.

\citet{Ryan99} conducted 20 low speed (100 m s$^{-1}$) impact
experiments into solid and porous ice targets.  They found that porous
ice targets, though significantly weaker than the solid ice targets
under static conditions, had a disruption strength equivalent to the
solid targets with similar total mass. The authors attribute this
behavior to the efficient dissipation of energy in void spaces.

\citet{Love93} ran a series of hypervelocity experiments (4.8-6.0
km~s$^{-1}$) into glass targets of varying porosity and strength. They
found that the specific energy needed to catastrophically disrupt the
target was proportional to (1-porosity)$^{-3.6}$. Their results
suggest that the porosity of the target is more important for the
collision outcome than the compressive strength of the target. More
work is needed to separate the effects of porosity and strength.
Impacts into the more porous targets resulted in deeper penetration of
the projectile but the excavated volume was about the same as in less
porous targets.  With higher porosity, the damage from the impact was
more localized.  These results suggest that porous objects in the
solar system would have longer lifetimes against collisional
disruption than monoliths of the same material.

\citet{Arakawa02} performed moderate speed impact experiments (150-670
m s$^{-1}$) into ice and ice-silicate mixtures to quantify the effect
of porosity on disruption strength. In pure ice targets, the
disruption strength increased with increasing porosity.  Puzzlingly,
in mixed material targets, the disruption strength decreased with
increasing porosity. These experiments suggest that the nature of the
material bonding (and material strength) can be as important as the
bulk porosity. 

The work to date demonstrates that porosity plays a significant role
in the outcome of catastrophic disruption experiments. However, more
work is needed to understand how porosity strengthens a material and
how to predict disruption strength as a function of porosity.

\subsubsection{Catastrophic Disruption Numerical Simulations}\label{sec:CDnum}

%{\bf CD modeling of macroporosity vs. contact monoliths: Asphaug --
 % contact binary case}
Investigations of catastrophic disruption in the gravity regime rely
upon numerical experiments. Studies including KBO analog materials are
limited.  \citet{Asphaug98} have considered km s$^{-1}$ impacts into
macroporous targets. \citet{Benz99} and {\em Leinhardt and Stewart, in prep.} have
studied the disruption of solid ice targets. More complex simulations
of KBOs including microporosity and mixed silicates with ice have yet
to be conducted.

Using a SPH code, \citet{Asphaug98} investigated how different
internal configurations affect the collision outcome.  They considered
5 km s$^{-1}$ rocky impacts onto a target shaped like asteroid
Castalia, which appears to be a contact binary. The possible internal
structures considered were (i) a solid rock, (ii) a global rubble pile
with 50\% bulk porosity, and (iii) two solid rock pieces separated by
a zone of highly damaged rock. In all three cases, the mass of the
target was constant (the density of the rock was changed). The model
included material strength and self-gravity.

In the rubble pile case, some of the energy generated by the impact is
partitioned into collapsing void space. In addition, the shock wave
reflects off of the free surfaces of the rubble pieces.  As a result,
shock effects were focused close to the impact site and the shock
pressures were dissipated much more quickly compared to the solid rock
target, in agreement with laboratory impact experiments (\S
\ref{sec:CDlab}). The velocities of the ejected material were higher
in the rubble pile configuration than the solid rock case, resulting
in a small ejecta blanket or none at all. In the two solid piece
model, the damaged region in the middle of the body reflects the shock
wave so that the damage is localized to the piece that was impacted.
This study elegantly demonstrates the importance of internal structure
in the outcome of collision events.

\begin{figure}
\includegraphics[width=80mm]{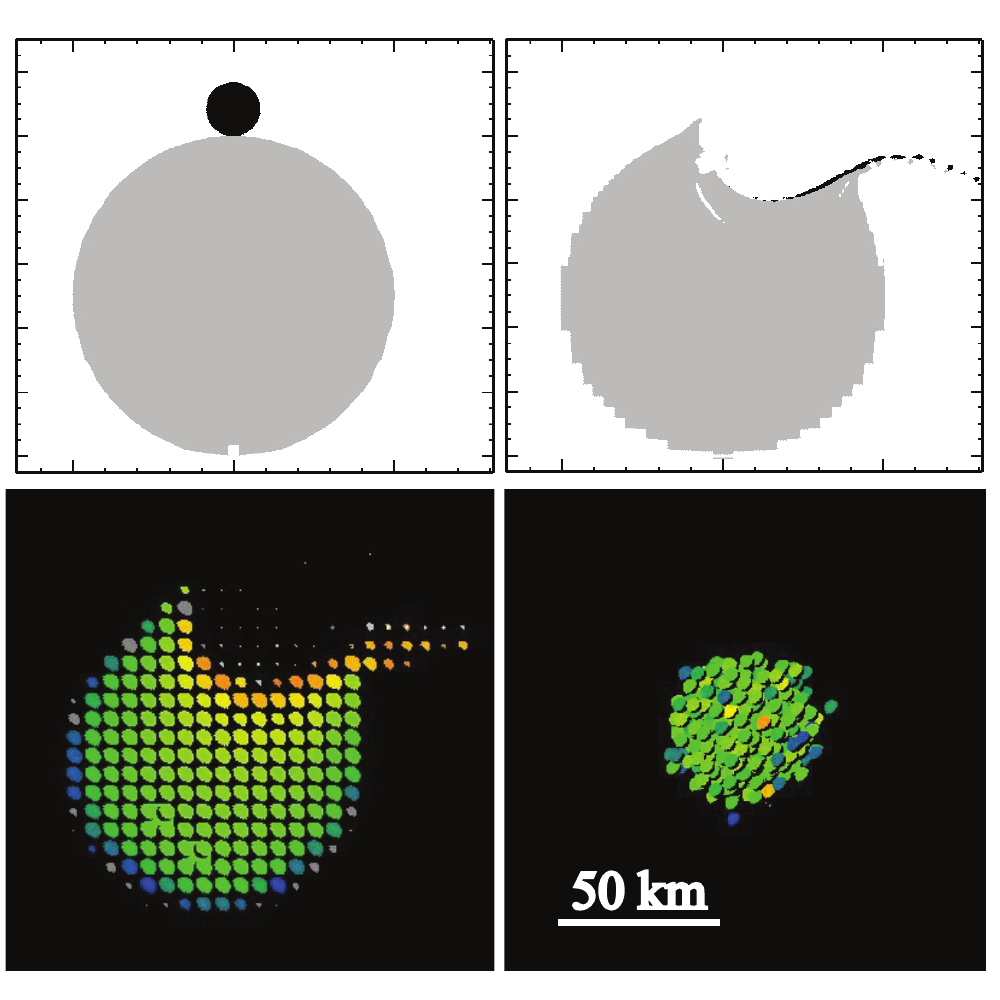} 
\caption{\small An example of a hybrid hydrocode to $N$-body
  numerical simulation of the catastrophic disruption of a solid ice
  target. The first row shows the positions of the target (grey) and
  projectile (black) at time 0 and 30 s. The second row shows the
  target after handoff from the hydrocode to the $N$-body code at 60 s
  and the largest remnant (r = 35 km) at 140 hr.  The projectile (r =
  8.4 km) hit the target (r = 50 km) at 45 degrees and 3 km s$^{-1}$.
  The first three frames show a slice through the 3D target and
  projectile along the y=0 plane. The final frame shows the surface of
  the largest post-collision remnant. The color coding in the $N$-body
  frames show the peak pressure attained by each mass element. Blue is
  the lowest peak pressure ($2\times 10^6$ dynes cm$^{-2}$), and red
  is the highest ($1\times 10^{11}$ dynes cm$^{-2}$). The peak
  pressure is stored in Lagrangian tracer particles during the
  hydrocode component of the simulation. The few non-colored (grey or
  white) particles in the $N$-body images are blocks from the
  hydrocode grid that did not contain any tracer particles. The
  surface of the largest remnant shows a mixture of high and low
  levels of shock deformation.
  \label{fig:strip}}
\end{figure}
 
A significant problem limiting numerical studies of hypervelocity
catastrophic disruption events is the vast difference in dynamical
times between the shock propagation and gravitational reaccumulation.
To make the problem more tractable, {\em Leinhardt and Stewart, in prep.} have begun
using a hybidized shock physics -- gravity method to study KBO analog
objects (Fig.~\ref{fig:strip}). The impact and deformation stage is
modeled using a shock physics code, {\tt CTH} \citep{McGlaun90}, and
the results are handed to a $N$-body gravity code, {\tt pkdgrav}
\citep{Stadel01,Richardson00,Leinhardt00}. This method allows detailed
modeling of the shock deformation including heating, phase changes,
and mixing of material as well as the final gravitational
reaccumulation of fragments.

These simulations record the provenance of the material in the largest
remnants and track the degree to which the reaccumulated material is
processed by the initial impact event.  For example, the peak shock
pressure (and hence the amount of melting or vaporization) experienced
by each mass element is recorded. In a catastrophic impact, a large
fraction of the original surface is lost, and the original surface of the
target is only maintained at the antipode of the largest
post-collision remnant. The surface materials on the largest remnant
reflect heterogeneous shock processing (Fig.~\ref{fig:strip}). Both
highly and weakly shocked material lines the surface, while the
interior material has a more homogeneous history of moderate shock
levels. This suggests that the surface materials on KBOs that have
suffered a catastrophic impacts could be heterogeneously devolatilized
in comparison to the interior.  The surface heterogeneity may also
lead to color variations.

%Future numerical work should investigate the effect of various mixed
%internal configurations of ice, basalt, and micro- and macro-porosity
%on collision outcome with a primary goal to determine the level of
%devolatilization from collisional evolution. These simulations will
%show how small bodies in our solar system evolve and to help explain
%the diversity of objects in the Kuiper Belt.

%Cratering and catastrophic disruption experiments have shown that
%taking into account all likely important characteristics such as
%mixtures of refractory and volatile elements, porosity both micro
%(compaction) and macro (rubble piles), and layering is very difficult
%since all characteristics that have been investigated seem to effect
%the collision outcome. It is not yet clear which characteristics if
%any dominate in determining the collision outcome for cratering
%impacts on KBOs. For catastrophic disruption impacts it appears that
%porosity is the most important in determining the collision outcome.
%But we do not yet have a complete model of shock propagation through a
%porous media and there has not been nearly as extensive investigation
%of the composition parameters of KBOs in the catastrophic disruption
%regime.

\section{Summary and Future Directions}\label{sec:questions}

In this chapter, we have discussed the present state of knowledge
about the possible physical effects of collisions in the Kuiper Belt.
The body of work on impact cratering and disruption events in KBO
analog materials demonstrate that composition and internal structure
(particularly porosity) have significant affects on the final outcome
of collision events. Understanding the role of collisions in changing
the composition and structure of KBOs is important because KBOs are
the best representatives of the planetesimals that accreted into the
outer solar system planets.

%The collisional evolution of KBOs is a difficult problem because there
%are several comet-like characteristics such as volatile content,
%porosity, layering, and mixing of materials that produce a broad range
%of collision outcomes. Although there has been great progress over the
%past two decades on understanding the dynamical history of the Kuiper
%Belt and a growing body of data on the physical properties of KBOs and
%progress has been made in understanding the physics governing
%collisions in icy, porous bodies models are still incomplete.

The range of possible outcomes of collisions in the Kuiper belt region
is more complicated than in the asteroid belt.  In particular, the low
relative impact velocities, the low mean density of KBOs, and the
likely presence of a variety of internal structures, are not fully
accounted for in present impact models.  Impact cratering scaling laws
and catastrophic disruption criteria that have been developed for
hypervelocity collisions on solid planetary surfaces and within the
asteroid belt may not be widely applicable to KBOs. For example, the
unanticipated large amount of mass ejected from comet Temple 1 from
the Deep Impact mission revealed that important physical processes are
missing from the crater scaling laws \citep[][]{Holsapple07}.

%-- add question of rise time, length scales to criteria for
 %  melting/vaporization; need for study of other volatiles.
  
%We will end this chapter with a list of outstanding questions and what
%we believe is necessary to answer these questions.

%\section{OUTSTANDING QUESTIONS}\label{sec:questions}

%Most previous studies did not encompass the full complexity of
%collisions in the Kuiper Belt. The development of more sophisticated
%simulations is an active area of research.
%The state of the art in the study of collisions in the Kuiper Belt has
%extended beyond the purely dynamical regime (e.g., evolution of the
%population size distribution). 

%The next step is to address questions
%on the evolution of the composition and structure of the bodies
%themselves, including:

We close with recommendations for areas of future work to advance our
knowledge of the properties of Kuiper Belt Objects:

\begin{enumerate}

\item What is the role of porosity in the outcome of cratering and
  catastrophic disruption collisions?  And how do we separate the
  effects of porosity and strength?
  
  Predicting the outcome of collisions into porous materials of 
  various strengths requires an improved understanding of (i) energy
  coupling into the target and (ii) shock-induced damage (degradation
  of strength). As both are difficult to model accurately in codes,
  clever laboratory experiments that include direct measurements of 
  shock wave decay, damage, and final crater sizes, in targets that
  vary porosity and strength independently are necessary.  In
  cratering events, the residual strength of the damaged target
  ultimately determines the final crater shape and the transition 
  between strength and gravity-dominated regimes. For disruption
  events, laboratory results will need to be incorporated into
  numerical simulations.
  
\item What is the correct way to scale laboratory experiments on 
  porous materials to larger scales?
  
  Two length scales appear to dominate the problem: (i) the length
  scale of the porosity with respect to the deforming shock wave
  (microporosity vs.\ macroporosity) and (ii) the depth of energy 
  coupling of the projectile. Laboratory and numerical experiments can
  directly address the effects of varying each length scale as the
  problem size increases from laboratory targets to planets. More
  information about the scale of porosities in KBOs can be obtained 
  from studies of comet nuclei and inferences of rubble pile vs.\ 
  differentiated internal structures in large KBOs.
\item How should volatility be incorporated into scaling laws?
  
  Vapor generation (or release) from a collision event would affect
  the momentum of the flow of excavated or dispersed material.  This
  difficult problem needs more information on the actual composition 
  of KBOs.
  
\item How does differentiation or layering affect the catastrophic
  disruption threshold?
  
  The propagation of the impact shock wave through the target is
  influenced by a layered internal structure. This tractable problem 
  can be addressed through laboratory and numerical experiments of
  plausible internal configurations in KBOs.
  
\item How does the mass ratio of the colliding bodies change the
  catastrophic disruption criteria in the hypervelocity regime? 
  
  The shock pressure profile through the target depends on the size
  and velocity of the projectile. Laboratory and numerical experiments
  can directly address this problem for solid bodies. Solutions to 
  question 1 in this list are required for highly porous bodies.
  
\item How can we validate numerical simulations in the gravity regime?
  
  Crater scaling laws have been validated by high gravity 
  (centrifuge) experiments. In the study of highly porous and weak
  materials, experiments in vacuum and under low gravity are also
  needed.  Validation of catastrophic disruption simulations in the
  gravity regime will require new techniques. 
  
\item What is the magnitude of modifications of KBOs from mutual
  collisions compared to other ``weathering'' processes? How different
  are present day KBOs from the primordial planetesimals in the outer 
  solar system?
  
  Cumulative changes in observable properties of KBOs, including
  densities, colors, composition, and internal structures, can be
  addressed by updating collisional evolution models of the Kuiper 
  Belt with the latest understanding of collisional processes in
  porous, icy bodies. Given the wide range of possible physical
  properties of KBOs, studies of individual collisions are warranted
  to examine common impact scenarios. At present there is no certain 
  answer, and our understanding will be driven by observations to
  come.
\end{enumerate}

\acknowledgements

\noindent{\bf ACKNOWLEDGEMENTS}

\vspace{.1cm}
The authors thank K. Housen, S. Kenyon, and E. Asphaug for careful review of this manuscript. We would also like to thank A. Morbidelli, L. Senft, M. Holman, J.-L. Margot, M. Brown, E. Schaller, and D. Raggozine for helpful discussions.
 
\bibliography{chapter}

\begin{thebibliography}{116}
\providecommand{\natexlab}[1]{#1}

\bibitem[{\emph{Ahrens and Johnson}(1995{\natexlab{a}})}]{ahrens95a}
Ahrens T. J. and Johnson M. L. (1995{\natexlab{a}}) Shock Wave Data for
  Minerals.
\newblock In \emph{Mineral Physics and Crystallography, A
  Handbook of Physical Constants}, vol.~2 (T. J. Ahrens ed.) pp.~143--184. Amer. Geophys. Union., Washington, D.C.

\bibitem[{\emph{Ahrens and Johnson}(1995{\natexlab{b}})}]{ahrens95b}
Ahrens T. J. and Johnson M. L. (1995{\natexlab{b}}) Shock Wave Data for Rocks.
\newblock In \emph{Rock Physics and Phase Relations, A
  Handbook of Physical Constants}, vol.~3 (T. J. Ahrens ed.), pp.~35--44. Amer. Geophys. Union, Washington, D.C.

\bibitem[{\emph{{Ahrens} and {Okeefe}}(1977)}]{Ahrens77}
{Ahrens} T. J. and {Okeefe} J. D. (1977) {Equations of State and Impact-Induced
  Shock-Wave Attenuation on the Moon}.
\newblock In \emph{Impact and Explosion Cratering: Planetary and Terrestrial
  Implications} (D. J. {Roddy}, R. O. {Pepin} and R. B. {Merrill}, eds.), pp. 639--656, Pergamon Press, Inc., New York.

\bibitem[{\emph{Ahrens and O'Keefe}(1987)}]{ahrens87}
Ahrens T. J. and O'Keefe J. D. (1987) Impact on the Earth, Ocean and
  Atmosphere.
\newblock \emph{International Journal of Impact Engineering}, \emph{5}, 13--32.

\bibitem[{\emph{{Arakawa} et~al.}(2000)\emph{{Arakawa}, {Higa},
  {Leliwa-Kopysty{\'n}ski} and {Maeno}}}]{Arakawa00}
{Arakawa} M., {Higa} M., {Leliwa-Kopysty{\'n}ski} J. and {Maeno} N. (2000)
  {Impact Cratering of Granular Mixture Targets made of H$_{2}$O Ice-CO$_{2}$
  Ice-Pyrophylite}.
\newblock \emph{\planss}, \emph{48}, 1437--1446.

\bibitem[{\emph{{Arakawa} et~al.}(2002)\emph{{Arakawa}, {Leliwa-Kopystynski}
  and {Maeno}}}]{Arakawa02}
{Arakawa} M., {Leliwa-Kopystynski} J. and {Maeno} N. (2002) {Impact Experiments
  on Porous Icy-Silicate Cylindrical Blocks and the Implication for Disruption
  and Accumulation of Small Icy Bodies}.
\newblock \emph{Icarus}, \emph{158}, 516--531.

\bibitem[{\emph{{Asphaug} et~al.}(1998)\emph{{Asphaug}, {Ostro}, {Hudson},
  {Scheeres} and {Benz}}}]{Asphaug98}
{Asphaug} E., {Ostro} S. J., {Hudson} R. S., {Scheeres} D. J. and {Benz} W.
  (1998) {Disruption of Kilometre-Sized Asteroids by Energetic Collisions}.
\newblock \emph{\nat}, \emph{393}, 437--440.

\bibitem[{\emph{{Asphaug} et~al.}(2002)\emph{{Asphaug}, {Ryan} and
  {Zuber}}}]{Asphaug02}
{Asphaug} E., {Ryan} E. V. and {Zuber} M. T. (2002) {Asteroid Interiors}.
\newblock In \emph{Asteroids III} (W. F. {Bottke}, A.~{Cellino}, P.~{Paolicchi} and R.~{Binzel}, eds.), pp. 463--484, Univ. of Arizona Press, Tuscon.

\bibitem[{\emph{{Barucci} et~al.}(2005)\emph{{Barucci}, {Cruikshank}, {Dotto},
  {Merlin}, {Poulet}, {Dalle Ore}, {Fornasier} and {de Bergh}}}]{Barucci05}
{Barucci} M. A., {Cruikshank} D. P., {Dotto} E., {Merlin} F., {Poulet} F.,
  {Dalle Ore} C., {Fornasier} S. and {de Bergh} C. (2005) {Is Sedna another
  Triton?}
\newblock \emph{\aap}, \emph{439}, L1--L4.

\bibitem[{\emph{{Belton} and {A'Hearn}}(1999)}]{Belton99}
{Belton} M. J. S. and {A'Hearn} M. F. (1999) {Deep Sub-Surface Exploration of
  Cometary Nuclei}.
\newblock \emph{Advances in Space Research}, \emph{24}, 1175--1183.

\bibitem[{\emph{Belton et~al.}(2007)\emph{Belton, Thomas, Veverka, Schultz,
  A'Hearn, Feaga, Farnham, Groussin, Li, Lisse, McFadden, Sunshine, Meech,
  Delamere and Kissel}}]{Belton07}
Belton M. J. S., Thomas P., Veverka J., Schultz P., A'Hearn M. F., Feaga L.,
  Farnham T., Groussin O., Li J. Y., Lisse C., McFadden L., Sunshine J., Meech
  K. J., Delamere W. A. and Kissel J. (2007) The Internal Structure of Jupiter
  Family Cometary Nuclei from Deep Impact Observations: The ``talps'' or
  ``layered pile'' Model.
\newblock \emph{Icarus}, \emph{187}, 1, 332.

\bibitem[{\emph{{Bendjoya} and {Zappal\`a}}(2002)}]{Bendjoya02}
{Bendjoya} P. and {Zappal\`a} V. (2002) Asteroid Family Identification.
\newblock In \emph{Asteroids III} (W. F. {Bottke}, A.~{Cellino}, P.~{Paolicchi} and R.~{Binzel}, eds.), pp. 613--618, Univ. of Arizona Press, Tuscon.

\bibitem[{\emph{{Benz} and {Asphaug}}(1999)}]{Benz99}
{Benz} W. and {Asphaug} E. (1999) {Catastrophic Disruptions Revisited}.
\newblock \emph{Icarus}, \emph{142}, 5--20.

\bibitem[{\emph{{Bernstein} et~al.}(2004)\emph{{Bernstein}, {Trilling},
  {Allen}, {Brown}, {Holman} and {Malhotra}}}]{Bernstein04}
{Bernstein} G. M., {Trilling} D. E., {Allen} R. L., {Brown} M. E., {Holman} M.
  and {Malhotra} R. (2004) {The Size Distribution of Trans-Neptunian Bodies}.
\newblock \emph{\aj}, \emph{128}, 1364--1390.

\bibitem[{\emph{{Brown} et~al.}(2007{\natexlab{a}})\emph{{Brown}, {Barkume},
  {Blake}, {Schaller}, {Rabinowitz}, {Roe} and {Trujillo}}}]{Brown06}
{Brown} M. E., {Barkume} K. M., {Blake} G. A., {Schaller} E. L., {Rabinowitz}
  D. L., {Roe} H. G. and {Trujillo} C. A. (2007{\natexlab{a}}) {Methane and
  Ethane on the Bright Kuiper Belt Object 2005 FY9}.
\newblock \emph{\aj}, \emph{133}, 284--289.

\bibitem[{\emph{{Brown} et~al.}(2007{\natexlab{b}})\emph{{Brown}, {Barkume},
  {Ragozzine} and {Schaller}}}]{Brown07}
{Brown} M. E., {Barkume} K. M., {Ragozzine} D. and {Schaller} E. L.
  (2007{\natexlab{b}}) {A Collisional Family of Icy Objects in the Kuiper
  Belt}.
\newblock \emph{Nature}, \emph{446}, 231--346.

\bibitem[{\emph{{Burchell} et~al.}(1998)\emph{{Burchell}, {Brooke-Thomas},
  {Leliwa-Kopystynski} and {Zarnecki}}}]{Burchell98}
{Burchell} M. J., {Brooke-Thomas} W., {Leliwa-Kopystynski} J. and {Zarnecki} J.
  C. (1998) {Hypervelocity Impact Experiments on Solid CO$_2$ Targets}.
\newblock \emph{Icarus}, \emph{131}, 210--222.

\bibitem[{\emph{{Burchell} and {Johnson}}(2005)}]{Burchell05}
{Burchell} M. J. and {Johnson} E. (2005) {Impact Craters on Small Icy Bodies
  such as Icy Satellites and Comet Nuclei}.
\newblock \emph{\mnras}, \emph{360}, 769--781.

\bibitem[{\emph{{Burchell} et~al.}(2002)\emph{{Burchell}, {Johnson} and
  {Grey}}}]{Burchell02}
{Burchell} M. J., {Johnson} E. and {Grey} I. D. S. (2002) {Hypervelocity
  impacts on porous ices}.
\newblock In \emph{ESA SP-500: Asteroids, Comets, and
  Meteors: ACM 2002} (B.~{Warmbein}, ed.), pp. 859--862, ESA Publications Division, Noordwijk, Netherlands.

\bibitem[{\emph{{Chapman} and {McKinnon}}(1986)}]{Chapman86}
{Chapman} C. R. and {McKinnon} W. B. (1986) {Cratering of Planetary
  Satellites}.
\newblock In \emph{Satellites} (J. A. Burns and M. S. Matthews, eds.), pp. 492--580, Univ. of Arizona Press, Tuscon.

\bibitem[{\emph{{Chen} et~al.}(2006)\emph{{Chen}, {Alcock}, {Axelrod},
  {Bianco}, {Byun}, {Chang}, {Cook}, {Dave}, {Giammarco}, {Kim}, {King}, {Lee},
  {Lehner}, {Lin}, {Lin}, {Lissauer}, {Marshall}, {Meinshausen}, {Mondal}, {de
  Pater}, {Porrata}, {Rice}, {Schwamb}, {Wang}, {Wang}, {Wen} and
  {Zhang}}}]{Chen06}
{Chen} W. P., {Alcock} C., {Axelrod} T., {Bianco} F. B., {Byun} Y. I., {Chang}
  Y. H., {Cook} K. H., {Dave} R., {Giammarco} J., {Kim} D. W., {King} S. K.,
  {Lee} T., {Lehner} M., {Lin} C. C., {Lin} H. C., {Lissauer} J. J., {Marshall}
  S., {Meinshausen} N., {Mondal} S., {de Pater} I., {Porrata} R., {Rice} J.,
  {Schwamb} M. E., {Wang} A., {Wang} S. Y., {Wen} C. Y. and {Zhang} Z. W.
  (2006) {Search for Small Trans-Neptunian Objects by the TAOS Project}.
\newblock In \emph{Proceedings IAU Symposium No. 236, 2007} (A. Milani, G. B. Valsecchi, and D. Vokrouhlicky, eds.), Intern. Astron. Union.

\bibitem[{\emph{{Cintala} et~al.}(1985)\emph{{Cintala}, {Smrekar}, {Horz} and
  {Cardenas}}}]{Cintala85}
{Cintala} M. J., {Smrekar} S., {Horz} F. and {Cardenas} F. (1985) {Impact
  Experiments in H$_2$O Ice, I: Cratering}.
\newblock In \emph{Lunar and Planetary Institute Conference Abstracts}, pp.
  131--132.

\bibitem[{\emph{{Croft} et~al.}(1979)\emph{{Croft}, {Kieffer} and
  {Ahrens}}}]{Croft79}
{Croft} S. K., {Kieffer} S. W. and {Ahrens} T. J. (1979) {Low-Velocity Impact
  Craters in Ice and Ice-Saturated Sand with Implications for Martian Crater
  Count Ages}.
\newblock \emph{\jgr}, \emph{84}, 8023--8032.

\bibitem[{\emph{{Davis} et~al.}(1979)\emph{{Davis}, {Chapman}, {Greenberg},
  {Weidenschilling} and {Harris}}}]{Davis79}
{Davis} D. R., {Chapman} C. R., {Greenberg} R., {Weidenschilling} S. J. and
  {Harris} A. W. (1979) {Collisional Evolution of Asteroids - Populations,
  Rotations, and Velocities}.
\newblock In \emph{Asteroids} (T.~Gehrels, ed.), pp. 528--557, Univ. of Arizona Press, Tuscon.

\bibitem[{\emph{{Davis} and {Farinella}}(1997)}]{Davis97}
{Davis} D. R. and {Farinella} P. (1997) {Collisional Evolution of
  Edgeworth-Kuiper Belt Objects}.
\newblock \emph{Icarus}, \emph{125}, 50--60.

\bibitem[{\emph{{Dell'Oro} et~al.}(2001)\emph{{Dell'Oro}, {Marzari},
  {Paolicchi} and {Vanzani}}}]{delloro01}
{Dell'Oro} A., {Marzari} F., {Paolicchi} P. and {Vanzani} V. (2001) Updated
  Collisional Probabilities of Minor Body Populations.
\newblock \emph{\aap}, \emph{366}, 1053--1060.

\bibitem[{\emph{{dell'Oro} et~al.}(2001)\emph{{dell'Oro}, {Paolicchi},
  {Cellino}, {Zappal{\`a}}, {Tanga} and {Michel}}}]{Delloro01b}
{dell'Oro} A., {Paolicchi} P., {Cellino} A., {Zappal{\`a}} V., {Tanga} P. and
  {Michel} P. (2001) {The Role of Families in Determining Collision Probability
  in the Asteroid Main Belt}.
\newblock \emph{Icarus}, \emph{153}, 52--60.

\bibitem[{\emph{{Dohnanyi}}(1969)}]{Dohnanyi69}
{Dohnanyi} J. W. (1969) {Collisional Models of Asteroids and their Debris}.
\newblock \emph{\jgr}, \emph{74}, 2531--2554.

\bibitem[{\emph{{Durda} et~al.}(2003)\emph{{Durda}, {Flynn} and {van
  Veghten}}}]{Durda03}
{Durda} D. D., {Flynn} G. J. and {van Veghten} T. W. (2003) {Impacts into
  Porous Foam Targets: Possible Implications for the Disruption of Comet
  Nuclei}.
\newblock \emph{Icarus}, \emph{163}, 504--507.

\bibitem[{\emph{Fujiwara et~al.}(2006)\emph{Fujiwara, Kawaguchi, Yeomans, Abe,
  Mukai, Okada, Saito, Yano, Yoshikawa, Scheeres, Barnouin-Jha, Cheng, Demura,
  Gaskell, Hirata, Ikeda, Kominato, Miyamoto, Nakamura, Nakamura, Sasaki and
  Uesugi}}]{Fujiwara06}
Fujiwara A., Kawaguchi J., Yeomans D. K., Abe M., Mukai T., Okada T., Saito J.,
  Yano H., Yoshikawa M., Scheeres D. J., Barnouin-Jha O., Cheng A. F., Demura
  H., Gaskell R. W., Hirata N., Ikeda H., Kominato T., Miyamoto H., Nakamura A.
  M., Nakamura R., Sasaki S. and Uesugi K. (2006) {The Rubble-Pile Asteroid
  Itokawa as Observed by Hayabusa}.
\newblock \emph{Science}, \emph{312}, 5778, 1330--1334.

\bibitem[{\emph{{Furnish} and {Remo}}(1997)}]{Furnish97}
{Furnish} M. D. and {Remo} J. L. (1997) {Ice Issues, Porosity, and Snow
  Experiments for Dynamic NEO and Comet Medeling}.
\newblock In \emph{{Near-Earth Objects: The United Nations International Conference}} (J. L. Remo, ed.), pp. 566--582, New York Academy of Science, New York.

\bibitem[{\emph{{Grey} and {Burchell}}(2003)}]{Grey03}
{Grey} I. D. S. and {Burchell} M. J. (2003) {Hypervelocity Impact Cratering on
  Water Ice Targets at Temperatures Ranging from 100 K to 253 K}.
\newblock \emph{J. Geophys. Res.}, \emph{108}, 6--1.

\bibitem[{\emph{Gross et~al.}(1988)\emph{Gross, Reichenauer and
  Fricke}}]{Gross88}
Gross J., Reichenauer G. and Fricke J. (1988) Mechanical-Properties of SiO$_2$
  Aerogels.
\newblock \emph{J. Phys. D}, \emph{21}, 9,
  1447--1451.

\bibitem[{\emph{{Gurnett} et~al.}(1997)\emph{{Gurnett}, {Ansher}, {Kurth} and
  {Granroth}}}]{Gurnett97}
{Gurnett} D. A., {Ansher} J. A., {Kurth} W. S. and {Granroth} L. J. (1997)
  {Micron-Sized Dust Particles Detected in the Outer Solar System by the
  Voyager 1 and 2 Plasma Wave Instruments}.
\newblock \emph{\grl}, \emph{24}, 3125--3128.

\bibitem[{\emph{{Hahn} and {Malhotra}}(1999)}]{Hahn99}
{Hahn} J. M. and {Malhotra} R. (1999) {Orbital Evolution of Planets Embedded in
  a Planetesimal Disk}.
\newblock \emph{\aj}, \emph{117}, 3041--3053.

\bibitem[{\emph{Herrmann}(1969)}]{Hermann1969}
Herrmann W. (1969) Constitutive Equation for the Dynamic Compaction of Ductile
  Porous Materials.
\newblock \emph{J. Applied Physics}, \emph{40}, 6, 2490--2499.

\bibitem[{\emph{{Holsapple} et~al.}(2002)\emph{{Holsapple}, {Giblin}, {Housen},
  {Nakamura} and {Ryan}}}]{Holsapple02}
{Holsapple} K., {Giblin} I., {Housen} K., {Nakamura} A. and {Ryan} E. (2002)
  {Asteroid Impacts: Laboratory Experiments and Scaling Laws}.
\newblock In \emph{Asteroids III} (W.F. {Bottke}, A.~{Cellino}, P.~{Paolicchi} and R.~{Binzel}, eds.), pp. 443--462, Univ. of Arizona Press, Tuscon.

\bibitem[{\emph{{Holsapple}}(1987)}]{Holsapple87b}
{Holsapple} K. A. (1987) {The Scaling of Impact Phenomenon}.
\newblock \emph{Int. J. Impact Eng.}, \emph{5}, 343--355.

\bibitem[{\emph{{Holsapple}}(1993)}]{Holsapple93}
{Holsapple} K. A. (1993) {The Scaling of Impact Processes in Planetary Sciences}.
\newblock \emph{Ann. Rev. Earth and Planet. Sci.}, \emph{21},
  333--373.

\bibitem[{\emph{Holsapple}(2007)}]{Holsapple07b}
Holsapple K. A. (2007) {Spin Limits of Solar System Bodies: From the Small
  Fast-Rotators to 2003 EL61}.
\newblock \emph{Icarus}, \emph{187}, 500--509.

\bibitem[{\emph{{Holsapple} and {Housen}}(2004)}]{Holsapple04}
{Holsapple} K. A. and {Housen} K. R. (2004) {The Cratering Database: Making
  Code Jockeys Honest}.
\newblock In \emph{Lunar and
  Planetary Institute Conference Abstracts}, no. 1779.

\bibitem[{\emph{Holsapple and Housen}(2007)}]{Holsapple07}
Holsapple K. A. and Housen K. R. (2007) A Crater and its Ejecta: An
  Interpretation of Deep Impact.
\newblock \emph{Icarus}, \emph{187}, 1, 345.

\bibitem[{\emph{{Holsapple} and {Schmidt}}(1987)}]{Holsapple87a}
{Holsapple} K. A. and {Schmidt} R. M. (1987) {Point Source Solutions and
  Coupling Parameters in Cratering Mechanics}.
\newblock \emph{\jgr}, \emph{92}, 6350--6376.

\bibitem[{\emph{Housen and Holsapple}(1999)}]{Housen99b}
Housen K. R. and Holsapple K. A. (1999) Scale Effects in Strength-Dominated
  Collisions of Rocky Asteroids.
\newblock \emph{Icarus}, \emph{142}, 1, 21--33.


\bibitem[{\emph{{Housen} and {Holsapple}}(2000)}]{Housen00}
{Housen} K. R. and {Holsapple} K. A. (2000) {Numerical Simulations of Impact
  Cratering in Porous Materials}.
\newblock In \emph{Lunar and Planetary Institute Conference Abstracts}, no. 1498.

\bibitem[{\emph{{Housen} and {Holsapple}}(2003)}]{Housen03}
{Housen} K. R. and {Holsapple} K. A. (2003) {Impact Cratering on Porous
  Asteroids}.
\newblock \emph{Icarus}, \emph{163}, 102--119.

\bibitem[{\emph{{Housen} et~al.}(1999)\emph{{Housen}, {Holsapple} and
  {Voss}}}]{Housen99}
{Housen} K. R., {Holsapple} K. A. and {Voss} M. E. (1999) {Compaction as the
  Origin of the Unusual Craters on the Asteroid Mathilde}.
\newblock \emph{\nat}, \emph{402}, 155--157.

\bibitem[{\emph{{Humes}}(1980)}]{Humes80}
{Humes} D. H. (1980) {Results of Pioneer 10 and 11 Meteoroid Experiments -
  Interplanetary and Near-Saturn}.
\newblock \emph{\jgr}, \emph{85}, 5841--5852.

\bibitem[{\emph{{Iijima} et~al.}(1995)\emph{{Iijima}, {Kato}, {Arakawa},
  {Maeno}, {Fujimura} and {Mizutani}}}]{Iijima95}
{Iijima} Y. i., {Kato} M., {Arakawa} M., {Maeno} N., {Fujimura} A. and
  {Mizutani} H. (1995) {Cratering Experiments on Ice: Dependence of Crater
  Formation on Projectile Materials and Scaling Parameter}.
\newblock \emph{\grl}, \emph{22}, 2005--2008.

\bibitem[{\emph{Jeanloz}(1989)}]{Jeanloz89}
Jeanloz R. (1989) Shock Wave Equation of State and Finite Strain Theory.
\newblock \emph{\jgr}, \emph{94}, B5, 5873--5886.

\bibitem[{\emph{{Jewitt} and {Luu}}(2000)}]{Jewitt00}
{Jewitt} D. C. and {Luu} J. X. (2000) Physical Nature of the Kuiper Belt.
\newblock In \emph{Protostars and Planets IV} (V. Mannings, A. P. Boss, and S. S. Russell, eds.), pp. 1201--1230, Univ. of Arizona Press, Tuscon.

\bibitem[{\emph{Johnson}(1991)}]{Johnson1991}
Johnson J. B. (1991) Simple Model of Shock-Wave Attenuation in Snow.
\newblock \emph{J. of Glaciology}, \emph{37}, 127, 303--312.

\bibitem[{\emph{{Kato} et~al.}(1995)\emph{{Kato}, {Iijima}, {Arakawa},
  {Okimura}, {Fujimura}, {Maeno} and {Mizutani}}}]{Kato95}
{Kato} M., {Iijima} Y., {Arakawa} M., {Okimura} Y., {Fujimura} A., {Maeno} N.
  and {Mizutani} H. (1995) {Ice-on-Ice Impact Experiments.}
\newblock \emph{Icarus}, \emph{113}, 423--441.

\bibitem[{\emph{{Kawakami} et~al.}(1983)\emph{{Kawakami}, {Mizutani}, {Takagi},
  {Kato} and {Kumazawa}}}]{Kawakami83}
{Kawakami} S., {Mizutani} H., {Takagi} Y., {Kato} M. and {Kumazawa} M. (1983)
  {Impact Experiments on Ice}.
\newblock \emph{\jgr}, \emph{88}, 5806--5814.

\bibitem[{\emph{{Kenyon} and {Bromley}}(2004)}]{Kenyon04}
{Kenyon} S. J. and {Bromley} B. C. (2004) {The Size Distribution of Kuiper Belt
  Objects}.
\newblock \emph{\aj}, \emph{128}, 1916--1926.

\bibitem[{\emph{{Kenyon} and {Luu}}(1999)}]{Kenyon99}
{Kenyon} S. J. and {Luu} J. X. (1999) {Accretion in the Early Kuiper Belt. II.
  Fragmentation}.
\newblock \emph{\aj}, \emph{118}, 1101--1119.

\bibitem[{\emph{{Koschny} and {Gr{\"u}n}}(2001)}]{Koschny01b}
{Koschny} D. and {Gr{\"u}n} E. (2001) {Impacts into Ice-Silicate Mixtures:
  Crater Morphologies, Volumes, Depth-to-Diameter Ratios, and Yield}.
\newblock \emph{Icarus}, \emph{154}, 391--401.

\bibitem[{\emph{{Koschny} et~al.}(2001)\emph{{Koschny}, {Kargl} and
  {Rott}}}]{Koschny01}
{Koschny} D., {Kargl} G. and {Rott} M. (2001) {Experimental Studies of the
  Cratering Process in Porous Ice Targets}.
\newblock \emph{Adv. Space Res.}, \emph{28}, 1533--1537.

\bibitem[{\emph{{Lacerda} and {Jewitt}}(2007)}]{Lacerda06}
{Lacerda} P. and {Jewitt} D. (2007) {Densities of Solar System Objects from
  their Rotational Lightcurves}.
\newblock \emph{\aj}, \emph{in press}.

\bibitem[{\emph{{Landgraf} et~al.}(2002)\emph{{Landgraf}, {Liou}, {Zook} and
  {Gr{\"u}n}}}]{Landgraf02}
{Landgraf} M., {Liou} J. C., {Zook} H. A. and {Gr{\"u}n} E. (2002) {Origins of
  Solar System Dust beyond Jupiter}.
\newblock \emph{\aj}, \emph{123}, 2857--2861.

\bibitem[{\emph{{Lange} and {Ahrens}}(1982)}]{Lange82}
{Lange} M. A. and {Ahrens} T. J. (1982) {Impact Cratering in - and Ice-Silicate
  Targets: an Experimental Assessment}.
\newblock In \emph{Lunar and Planetary Institute Conference Abstracts}, pp.
  415--416.

\bibitem[{\emph{{Lange} and {Ahrens}}(1983)}]{Lange83}
{Lange} M. A. and {Ahrens} T. J. (1983) {The Dynamic Tensile Strength of Ice
  and Ice Silicate Mixtures}.
\newblock \emph{\jgr}, \emph{88}, 1197--1208.

\bibitem[{\emph{{Lange} and {Ahrens}}(1987)}]{Lange87}
{Lange} M. A. and {Ahrens} T. J. (1987) {Impact Experiments in Low-Temperature
  Ice}.
\newblock \emph{Icarus}, \emph{69}, 506--518.

\bibitem[{\emph{{Leinhardt} and {Richardson}}(2002)}]{Leinhardt02}
{Leinhardt} Z. M. and {Richardson} D. C. (2002) {N-Body Simulations of
  Planetesimal Evolution: Effect of Varying Impactor Mass Ratio}.
\newblock \emph{Icarus}, \emph{159}, 306--313.

\bibitem[{\emph{{Leinhardt} et~al.}(2000)\emph{{Leinhardt}, {Richardson} and
  {Quinn}}}]{Leinhardt00}
{Leinhardt} Z. M., {Richardson} D. C. and {Quinn} T. (2000) {Direct N-body
  Simulations of Rubble Pile Collisions}.
\newblock \emph{Icarus}, \emph{146}, 133--151.

\bibitem[{\emph{{Love} and {Ahrens}}(1997)}]{Love97}
{Love} S. G. and {Ahrens} T. J. (1997) {Origin of Asteroid Rotation Rates in
  Catastrophic Impacts}.
\newblock \emph{\nat}, \emph{386}, 154--156.

\bibitem[{\emph{{Love} et~al.}(1993)\emph{{Love}, {H{\"o}rz} and
  {Brownlee}}}]{Love93}
{Love} S. G., {H{\"o}rz} F. and {Brownlee} D. E. (1993) {Target Porosity
  Effects in Impact Cratering and Collisional Disruption}.
\newblock \emph{Icarus}, \emph{105}, 216--224.

\bibitem[{\emph{{Margot}}(2002)}]{Margot02}
{Margot} J. L. (2002) {Astronomy: Worlds of Mutual Motion}.
\newblock \emph{\nat}, \emph{416}, 694--695.

\bibitem[{\emph{McGlaun et~al.}(1990)\emph{McGlaun, Thompson and
  Elrick}}]{McGlaun90}
McGlaun J. M., Thompson S. L. and Elrick M. G. (1990) CTH: A 3-Dimensional
  Shock-Wave Physics Code.
\newblock \emph{Int. J. Imp. Eng.}, \emph{10},
  351--360.

\bibitem[{\emph{{Mellor}}(1975)}]{Mellor75}
{Mellor} M. (1975) A Review of Basic Snow Mechanics.
\newblock In \emph{Snow Mechanics, Proceedings of the Grendelwald Symposium},
  pp. 251--291.

\bibitem[{\emph{{Melosh}}(1977)}]{Melosh77}
{Melosh} H. J. (1977) {Crater Modification by Gravity - A Mechanical Analysis
  of Slumping}.
\newblock In \emph{Impact and Explosion Cratering: Planetary and Terrestrial
  Implications} (D.J. {Roddy}, R.O. {Pepin} and R.B. {Merrill}, eds), pp. 1245--1260, Pergamon Press, Inc., New York.

\bibitem[{\emph{{Melosh}}(1989)}]{Melosh89}
{Melosh} H. J. (1989) \emph{{Impact Cratering}}.
\newblock {Oxford University Press, New York}.

\bibitem[{\emph{{Melosh} and {Ivanov}}(1999)}]{Melosh99}
{Melosh} H. J. and {Ivanov} B. A. (1999) {Impact Crater Collapse}.
\newblock \emph{Ann. Rev. Earth and Plan. Sci.}, \emph{27},
  385--415.

\bibitem[{\emph{{Melosh} and {McKinnon}}(1978)}]{Melosh78}
{Melosh} H. J. and {McKinnon} W. B. (1978) {The Mechanics of Ringed Basin
  Formation}.
\newblock \emph{\grl}, \emph{5}, 985--988.

\bibitem[{\emph{{Melosh} and {Ryan}}(1997)}]{Melosh97}
{Melosh} H. J. and {Ryan} E. V. (1997) {Asteroids: Shattered but Not
  Dispersed}.
\newblock \emph{Icarus}, \emph{129}, 562--564.

\bibitem[{\emph{{Merk} and {Prialnik}}(2006)}]{Merk06}
{Merk} R. and {Prialnik} D. (2006) {Combined Modeling of Thermal evolution and
  Accretion of Trans-Neptunian Objects - Occurrence of High Temperatures and
  Liquid Water}.
\newblock \emph{Icarus}, \emph{183}, 283--295.

\bibitem[{\emph{{Meyers}}(2001)}]{Meyers01}
{Meyers} M. A. (2001) \emph{{Dynamic Behavior of Materials}}.
\newblock {John Wiley and Sons, New York}.

\bibitem[{\emph{{M{\"u}ller} et~al.}(2005)\emph{{M{\"u}ller}, {{\'A}brah{\'a}m}
  and {Crovisier}}}]{Muller05}
{M{\"u}ller} T. G., {{\'A}brah{\'a}m} P. and {Crovisier} J. (2005) {Comets,
  Asteroids and Zodiacal Light as Seen by Iso}.
\newblock \emph{Space Sci. Rev.}, \emph{119}, 141--155.

\bibitem[{\emph{{Oberbeck} and {Quaide}}(1967)}]{Oberbeck67}
{Oberbeck} V. R. and {Quaide} W. L. (1967) {Estimated Thickness of a Fragmental
  Surface Layer of Oceanus Procellarum}.
\newblock \emph{J. Geophys. Res.}, \emph{72}, 4697--4704.

\bibitem[{\emph{O'Brien and Greenberg}(2003)}]{obrien03}
O'Brien D. P. and Greenberg R. (2003) Steady-State Size Distributions for
  Collisional Populations: Analytical Solution with Size-Dependent Strength.
\newblock \emph{Icarus}, \emph{164}, 2, 334.

\bibitem[{\emph{{O'Brien} and {Greenberg}}(2005)}]{Obrien05}
{O'Brien} D. P. and {Greenberg} R. (2005) {The Collisional and Synamical
  Evolution of the Main-Belt and NEA Size Distributions}.
\newblock \emph{Icarus}, \emph{178}, 179--212.

\bibitem[{\emph{{Pan} and {Sari}}(2005)}]{Pan05}
{Pan} M. and {Sari} R. (2005) {Shaping the Kuiper Belt Size Distribution by
  Shattering Large but Strengthless Bodies}.
\newblock \emph{Icarus}, \emph{173}, 342--348.

\bibitem[{\emph{{Paolicchi} et~al.}(2002)\emph{{Paolicchi}, {Burns} and
  {Weidenschilling}}}]{Paolicchi02}
{Paolicchi} P., {Burns} J. A. and {Weidenschilling} S. J. (2002) {Side Effects
  of Collisions: Spin Rate Changes, Tumbling Rotation States, and Binary
  Asteroids}.
\newblock In \emph{Asteroids III} (W.F. {Bottke}, A.~{Cellino}, P.~{Paolicchi} and R.~{Binzel},
  eds.), pp. 517--526, Univ. of Arizona Press, Tuscon.

\bibitem[{\emph{{Petit} et~al.}(2006)\emph{{Petit}, {Holman}, {Gladman},
  {Kavelaars}, {Scholl} and {Loredo}}}]{petit06}
{Petit} J. M., {Holman} M. J., {Gladman} B. J., {Kavelaars} J. J., {Scholl} H.
  and {Loredo} T. J. (2006) {The Kuiper Belt Luminosity Function from m$_{R}$=
  22 to 25}.
\newblock \emph{\mnras}, \emph{365}, 429--438.

\bibitem[{\emph{{Petrenko} and {Whitworth}}(1999)}]{Petrenko99}
{Petrenko} V. F. and {Whitworth} R. W. (1999) \emph{{The Physics of Ice}}.
\newblock Oxford University Press, New York.

\bibitem[{\emph{{Pierazzo} and {Melosh}}(2000)}]{Pierazzo00}
{Pierazzo} E. and {Melosh} H. J. (2000) {Melt Production in Oblique Impacts}.
\newblock \emph{Icarus}, \emph{145}, 252--261.

\bibitem[{\emph{{Pierazzo} et~al.}(1997)\emph{{Pierazzo}, {Vickery} and
  {Melosh}}}]{Pierazzo97}
{Pierazzo} E., {Vickery} A. M. and {Melosh} H. J. (1997) {A Reevaluation of
  Impact Melt Production}.
\newblock \emph{Icarus}, \emph{127}, 408--423.

\bibitem[{\emph{Poirier}(2000)}]{Poirier00}
Poirier J. P. (2000) \emph{{Introduction to the Physics of the Earth's
  Interior}}.
\newblock {Cambridge University Press, New York}.

\bibitem[{\emph{{Rabinowitz} et~al.}(2006)\emph{{Rabinowitz}, {Barkume},
  {Brown}, {Roe}, {Schwartz}, {Tourtellotte} and {Trujillo}}}]{Rabinowitz06}
{Rabinowitz} D. L., {Barkume} K., {Brown} M. E., {Roe} H., {Schwartz} M.,
  {Tourtellotte} S. and {Trujillo} C. (2006) {Photometric Observations
  Constraining the Size, Shape, and Albedo of 2003 EL61, a Rapidly Rotating,
  Pluto-sized Object in the Kuiper Belt}.
\newblock \emph{\apj}, \emph{639}, 1238--1251.

%\bibitem[{\emph{{Rice}}(1958)}]{Rice58}
%{Rice} D. A. (1958) \emph{{Gravity Control Measurements in North America.}}
%\newblock U.S.~Dept.~of Commerce, Coast and Geodetic Survey, Washington.

\bibitem[{\emph{{Rice}}(1958)}]{Rice58}
{Rice} M. H., McQueen, R. G., and Walsh, J. M. (1958) {Compression of Solids by Strong Shock Waves}, \emph{Solid State Physics}, \emph{6}, 1--63. 

\bibitem[{\emph{{Richardson} et~al.}(2000)\emph{{Richardson}, {Quinn}, {Stadel}
  and {Lake}}}]{Richardson00}
{Richardson} D. C., {Quinn} T., {Stadel} J. and {Lake} G. (2000) {Direct
  Large-Scale N-Body Simulations of Planetesimal Dynamics}.
\newblock \emph{Icarus}, \emph{143}, 45--59.

\bibitem[{\emph{{Roques} et~al.}(2006)\emph{{Roques}, {Doressoundiram},
  {Dhillon}, {Marsh}, {Bickerton}, {Kavelaars}, {Moncuquet}, {Auvergne},
  {Belskaya}, {Chevreton}, {Colas}, {Fernandez}, {Fitzsimmons}, {Lecacheux},
  {Mousis}, {Pau}, {Peixinho} and {Tozzi}}}]{Roques06}
{Roques} F., {Doressoundiram} A., {Dhillon} V., {Marsh} T., {Bickerton} S.,
  {Kavelaars} J. J., {Moncuquet} M., {Auvergne} M., {Belskaya} I., {Chevreton}
  M., {Colas} F., {Fernandez} A., {Fitzsimmons} A., {Lecacheux} J., {Mousis}
  O., {Pau} S., {Peixinho} N. and {Tozzi} G. P. (2006) {Exploration of the
  Kuiper Belt by High-Precision Photometric Stellar Occultations: First
  Results}.
\newblock \emph{\aj}, \emph{132}, 819--822.

\bibitem[{\emph{Ruoff}(1967)}]{Ruoff67}
Ruoff A. L. (1967) Linear Shock-Velocity-Particle-Velocity Relationship.
\newblock \emph{J. of Applied Physics}, \emph{38}, 13, 4976--4980.

\bibitem[{\emph{{Ryan} et~al.}(1999)\emph{{Ryan}, {Davis} and
  {Giblin}}}]{Ryan99}
{Ryan} E. V., {Davis} D. R. and {Giblin} I. (1999) {A Laboratory Impact Study
  of Simulated Edgeworth-Kuiper Belt Objects}.
\newblock \emph{Icarus}, \emph{142}, 56--62.

\bibitem[{\emph{{Ryan} et~al.}(1991)\emph{{Ryan}, {Hartmann} and
  {Davis}}}]{Ryan91}
{Ryan} E. V., {Hartmann} W. K. and {Davis} D. R. (1991) {Impact Experiments.
  III - Catastrophic Fragmentation of Aggregate Targets and Relation to
  Asteroids}.
\newblock \emph{Icarus}, \emph{94}, 283--298.

\bibitem[{\emph{Sammonds et~al.}(1998)\emph{Sammonds, Murrell and
  Rist}}]{Sammonds98}
Sammonds P. R., Murrell S. A. F. and Rist M. A. (1998) Fracture of Multiyear
  Sea Ice.
\newblock \emph{\jgr}, \emph{103}, C10,
  21795--21815.

\bibitem[{\emph{{Schmidt}}(1980)}]{Schmidt80}
{Schmidt} R. M. (1980) {Meteor Crater: Energy of Formation - Implications of
  Centrifuge Scaling}.
\newblock In \emph{Lunar and Planetary Science Conference}, pp. 2099--2128.

\bibitem[{\emph{{Schultz}}(1996)}]{Schultz96}
{Schultz} P. H. (1996) {Effect of Impact Angle on Vaporization}.
\newblock \emph{\jgr}, \emph{101}, 21117--21136.

\bibitem[{\emph{{Schultz}}(2003)}]{Schultz03}
{Schultz} P. H. (2003) {Impacts into Porous Volatile-Rich Substrates on Mars}.
\newblock In \emph{Sixth Intern. Conf. on Mars}, no. 3263.

\bibitem[{\emph{{Schultz} et~al.}(2005)\emph{{Schultz}, {Ernst} and
  {Anderson}}}]{Schultz05}
{Schultz} P. H., {Ernst} C. M. and {Anderson} J. L. B. (2005) {Expectations for
  Crater Size and Photometric Evolution from the Deep Impact Collision}.
\newblock \emph{Sp. Sci. Reviews}, \emph{117}, 207--239.

\bibitem[{\emph{{Schultz} and {Gault}}(1985)}]{Schultz85}
{Schultz} P. H. and {Gault} D. E. (1985) {Clustered Impacts - Experiments and
  Implications}.
\newblock \emph{\jgr}, \emph{90}, 3701--3732.

\bibitem[{\emph{{Senft} and {Stewart}}(2006)}]{Senft06}
{Senft} L. E. and {Stewart} S. T. (2006) {Modeling Impact Cratering into
  Layered Targets}.
\newblock \emph{\jgr}, \emph{submitted}.

\bibitem[{\emph{{Shrine} et~al.}(2002)\emph{{Shrine}, {Burchell} and
  {Grey}}}]{Shrine02}
{Shrine} N. R. G., {Burchell} M. J. and {Grey} I. D. S. (2002) {Velocity
  Scaling of Impact Craters in Water Ice over the Range 1 to 7.3 km s$^{-1}$}.
\newblock \emph{Icarus}, \emph{155}, 475--485.

\bibitem[{\emph{{Stadel}}(2001)}]{Stadel01}
{Stadel} J. G. (2001) {Cosmological N-body Simulations and their Analysis}.
\newblock \emph{Ph.D.~Thesis}.

\bibitem[{\emph{{Stern}}(1996)}]{Stern96}
{Stern} S. A. (1996) {On the Collisional Environment, Accretion Time Scales,
  and Architecture of the Massive, Primordial Kuiper Belt.}
\newblock \emph{\aj}, \emph{112}, 1203--1211.

\bibitem[{\emph{{Stern}}(2003)}]{Stern03}
{Stern} S. A. (2003) {The evolution of Comets in the Oort Cloud and Kuiper
  Belt}.
\newblock \emph{\nat}, \emph{424}, 639--642.

\bibitem[{\emph{{Stern} and {Colwell}}(1997)}]{Stern97}
{Stern} S. A. and {Colwell} J. E. (1997) {Collisional Erosion in the Primordial
  Edgeworth-Kuiper Belt and the Generation of the 30-50 AU Kuiper Gap}.
\newblock \emph{\apj}, \emph{490}, 879--882.

\bibitem[{\emph{{Stewart} and {Ahrens}}(2004)}]{Stewart04}
{Stewart} S. T. and {Ahrens} T. J. (2004) {A New H$_2$O Ice Hugoniot:
  Implications for Planetary Impact Events}.
\newblock In \emph{Shock Compression of
  Condensed Matter-2003} ({Furnish, M.~D.~et al.}, ed.), pp. 1478--1483. AIP.

\bibitem[{\emph{{Stewart} and {Ahrens}}(2005)}]{Stewart05}
{Stewart} S. T. and {Ahrens} T. J. (2005) {Shock Properties of H$_{2}$O Ice}.
\newblock \emph{\jgr}, \emph{110}, E9,
  3005.

\bibitem[{\emph{{Strom} et~al.}(2005)\emph{{Strom}, {Malhotra}, {Ito},
  {Yoshida} and {Kring}}}]{Strom05}
{Strom} R. G., {Malhotra} R., {Ito} T., {Yoshida} F. and {Kring} D. A. (2005)
  {The Origin of Planetary Impactors in the Inner Solar System}.
\newblock \emph{Science}, \emph{309}, 1847--1850.

\bibitem[{\emph{{Trujillo} et~al.}(2001)\emph{{Trujillo}, {Jewitt} and
  {Luu}}}]{Trujillo01}
{Trujillo} C. A., {Jewitt} D. C. and {Luu} J. X. (2001) {Properties of the
  Trans-Neptunian Belt: Statistics from the Canada-France-Hawaii Telescope
  Survey}.
\newblock \emph{\aj}, \emph{122}, 457--473.

\bibitem[{\emph{{Veverka} et~al.}(1997)\emph{{Veverka}, {Thomas}, {Harch},
  {Clark}, {Bell}, {Carcich}, {Joseph}, {Chapman}, {Merline}, {Robinson},
  {Malin}, {McFadden}, {Murchie}, {Hawkins}, {Farquhar}, {Izenberg} and
  {Cheng}}}]{Veverka97}
{Veverka} J., {Thomas} P., {Harch} A., {Clark} B., {Bell} III J. F., {Carcich}
  B., {Joseph} J., {Chapman} C., {Merline} W., {Robinson} M., {Malin} M.,
  {McFadden} L. A., {Murchie} S., {Hawkins} III S. E., {Farquhar} R.,
  {Izenberg} N. and {Cheng} A. (1997) {NEAR's Flyby of 253 Mathilde: Images of
  a C Asteroid}.
\newblock \emph{Science}, \emph{278}, 2109--2114.

\bibitem[{\emph{Weissman et~al.}(2005)\emph{Weissman, Asphaug and
  Lowry}}]{Weissman05}
Weissman P. R., Asphaug E. and Lowry S. C. (2005) 
  Structure and Density of Cometary Nuclei, In \emph{Comets II} (M. Festou, H. U. Keller, H. A. Weaver, eds.), pp. 337--357,
\newblock Univ. Arizona Press, Tuscon.

\bibitem[{\emph{Williams and Wetherill}(1994)}]{williams94}
Williams D. R. and Wetherill G. W. (1994) Size Distribution of Collisionally
  Evolved Asteroidal Populations: Analytical Solution for Self-Similar
  Collision Cascades.
\newblock \emph{Icarus}, \emph{107}, 1, 117.

\bibitem[{\emph{{W{\"u}nnemann} et~al.}(2006)\emph{{W{\"u}nnemann}, {Collins}
  and {Melosh}}}]{Wunnemann06}
{W{\"u}nnemann} K., {Collins} G. S. and {Melosh} H. J. (2006) {A Strain-Based
  Porosity Model for use in Hydrocode Simulations of Impacts and Implications
  for Transient Crater Growth in Porous Targets}.
\newblock \emph{Icarus}, \emph{180}, 514--527.

\end{thebibliography}
%PROCTOR TM LOW-TEMPERATURE SPEED OF SOUND IN SINGLE-CRYSTAL ICE 
%JOURNAL OF THE ACOUSTICAL SOCIETY OF AMERICA 39 (5P1): 972 \& 1966 

%Ice Issues, Porosity, and Snow Experiments for Dynamic NEO and Comet
%Modeling M. D. FURNISH and J. L. REMO Near-Earth Objects: The United
%Nations Conference Volume 822 published May 1997 Ann. N.Y. Acad. Sci.
%822: 566-582 (1997).

\end{document}